\documentclass[review]{elsarticle}

\usepackage{lineno,hyperref}
\modulolinenumbers[5]

\usepackage[capitalize]{cleveref}                        
\usepackage{siunitx}
\usepackage{subfig}
\usepackage{xspace}

\usepackage{tikz}
\newlength\imageheight
\usepackage{color}

\journal{Nucl. Instrum. Meth. Phys. Res. Sect. A}

\begin{document}
\begin{frontmatter}

\title{The tracking detector of the FASER experiment}

\author[1]{Henso Abreu}

\author[2]{Claire Antel}

\author[3,21]{Akitaka Ariga}

\author[4]{Tomoko Ariga}

\author[5]{Florian Bernlochner}

\author[5]{Tobias Boeckh}

\author[6]{Jamie Boyd}

\author[6]{Lydia Brenner}

\author[2]{Franck Cadoux}

\author[7]{David~W.~Casper}

\author[8]{Charlotte Cavanagh}

\author[9]{Xin Chen}

\author[11]{Andrea Coccaro}

\author[6]{Olivier Crespo-Lopez}

\author[12]{Sergey Dmitrievsky}

\author[8]{Monica D’Onofrio}

\author[9]{Candan Dozen}

\author[23]{Abdallah Ezzat}

\author[2]{Yannick Favre}

\author[13]{Deion Fellers}

\author[7]{Jonathan~L.~Feng}

\author[2]{Didier Ferrere}

\author[14]{Stephen Gibson}

\author[2]{Sergio Gonzalez-Sevilla}

\author[12]{Yuri Gornushkin}

\author[8]{Carl Gwilliam}

\author[15]{Shih-Chieh Hsu}

\author[9]{Zhen Hu}

\author[2]{Giuseppe Iacobucci}

\author[9]{Tomohiro Inada}

\author[6]{Sune Jakobsen}

\author[1]{Enrique Kajomovitz}

\author[16]{Felix Kling}

\author[6]{Umut Kose}

\author[6]{Susanne Kuehn}

\author[14]{Helena Lefebvre}

\author[17]{Lorne Levinson}

\author[15]{Ke Li}

\author[9]{Jinfeng Liu}

\author[2]{Chiara Magliocca}

\author[18]{Josh McFayden}

\author[2]{Matteo Milanesio}

\author[6]{Sam Meehan}

\author[6]{Dimitar Mladenov}

\author[2]{Théo Moretti}

\author[2]{Magdalena Munker}

\author[19]{Mitsuhiro Nakamura}

\author[19]{Toshiyuki Nakano}

\author[6]{Marzio Nessi}

\author[10]{Friedemann Neuhaus}

\author[14]{Laurie Nevay}

\author[4]{Hidetoshi Otono\corref{correspondingauthor}}
\cortext[correspondingauthor]{Corresponding author.}
\ead{otono@phys.kyushu-u.ac.jp}

%\author[7]{Johanna Paine}

\author[2]{Carlo Pandini}

\author[9]{Hao Pang}

\author[2]{Lorenzo Paolozzi}

\author[6]{Brian Petersen}

\author[6]{Francesco Pietropaolo}

\author[5]{Markus Prim}

\author[6]{Michaela Queitsch-Maitland}

\author[6]{Filippo Resnati}

\author[2]{Chiara Rizzi}

\author[19]{Hiroki Rokujo}

\author[10]{Elisa Ruiz-Choliz}

\author[6]{Jakob Salfeld-Nebgen}

\author[19]{Osamu Sato}

\author[3,22]{Paola Scampoli}

\author[10]{Kristof Schmieden}

\author[10]{Matthias Schott}

\author[2]{Anna Sfyrla}

\author[7]{Savannah Shively}

\author[15]{John Spencer}

\author[20]{Yosuke Takubo}

\author[2]{Noshin Tarannum}

\author[2]{Ondrej Theiner}

\author[13]{Eric Torrence}

\author[6]{Serhan Tufanli}

\author[12]{Svetlana Vasina}

\author[6]{Benedikt Vormwald}

\author[9]{Di Wang}

%\author[9]{Gang Zhang}

%%%%%%%%%%%%%%%%%%%%%%%%
\address[1]{Department of Physics and Astronomy, Technion---Israel Institute of Technology, Haifa 32000, Israel}

\address[2]{D\'epartement de Physique Nucl\'eaire et Corpusculaire, University of Geneva, CH-1211 Geneva 4, Switzerland}

%\address[3]{Universit\"at Bern, Sidlerstrasse 5, CH-3012 Bern, Switzerland}

\address[3]{Albert Einstein Center for Fundamental Physics, Laboratory for High Energy Physics, University of Bern, Sidlerstrasse 5, CH-3012 Bern, Switzerland}

\address[4]{Kyushu University, Nishi-ku, 819-0395 Fukuoka, Japan}

\address[5]{Universit\"at Bonn, Regina-Pacis-Weg 3, D-53113 Bonn, Germany}

\address[6]{CERN, CH-1211 Geneva 23, Switzerland}

\address[7]{Department of Physics and Astronomy, 
University of California, Irvine, CA 92697-4575, USA}

\address[8]{University of Liverpool, Liverpool L69 3BX, United Kingdom}

\address[9]{Department of Physics, Tsinghua University, Beijing, China}

\address[11]{INFN Sezione di Genova, Via Dodecaneso, 33--16146, Genova, Italy}

\address[12]{Joint Institute for Nuclear Research, Dubna, Russia}

\address[23]{Facult\'e de Physique et Ing\'enierie, University of Strasbourg, 3-5 rue de l’Universit\'e, Strasbourg, France, 67000}

\address[13]{University of Oregon, Eugene, OR 97403, USA}

\address[14]{Royal Holloway, University of London, Egham, TW20 0EX, UK}

\address[15]{Department of Physics, University of Washington, PO Box 351560, Seattle, WA 98195-1560, USA}

\address[16]{Deutsches Elektronen-Synchrotron DESY, Notkestr. 85, 22607 Hamburg, Germany}

\address[17]{Department of Particle Physics and Astrophysics, Weizmann Institute of Science, Rehovot 76100, Israel}

\address[18]{Department of Physics \& Astronomy, University of Sussex, Sussex House, Falmer, Brighton, BN1 9RH, United Kingdom}

\address[19]{Nagoya University, Furo-cho, Chikusa-ku, Nagoya 464-8602, Japan}

\address[10]{Institut f\"ur Physik, Universität Mainz, Mainz, Germany}

\address[20]{Institute of Particle and Nuclear Study, 
KEK, Oho 1-1, Tsukuba, Ibaraki 305-0801, Japan}

\address[21]{Department of Physics, Chiba University, 1-33 Yayoi-cho Inage-ku, Chiba, 263-8522, Japan}

\address[22]{Dipartimento di Fisica ``Ettore Pancini'', Universit\`a di Napoli Federico II, Complesso Universitario di Monte S. Angelo, I-80126 Napoli, Italy}





\begin{abstract}

FASER is a new experiment designed to search for new light weakly-interacting long-lived particles (LLPs) and study high-energy neutrino interactions in the very forward region of the LHC collisions at CERN. The experimental apparatus is situated 480~m downstream of the ATLAS interaction-point aligned with the beam collision axis. 
The FASER detector includes four identical tracker stations constructed from silicon microstrip detectors. Three of the tracker stations form a tracking spectrometer, and enable FASER to detect the decay products of LLPs decaying inside the apparatus, whereas the fourth station is used for the neutrino analysis. The spectrometer has been installed in the LHC complex since March 2021, while the fourth station is not yet installed. FASER will start physics data taking when the LHC resumes operation in early 2022. This paper describes the design, construction and testing of the tracking spectrometer, including the associated components such as the mechanics, readout electronics, power supplies and cooling system.
\end{abstract}

\begin{keyword}
Silicon microstrip detectors, Tracking detectors, FASER, LHC 
%\texttt{elsarticle.cls}\sep \LaTeX\sep Elsevier \sep template
%\MSC[2010] 00-01\sep  99-00
\end{keyword}

\end{frontmatter}

\vspace*{\fill}
\noindent
\copyright 2021 CERN for the benefit of the FASER Collaboration. Reproduction of this article or parts of it is allowed as specified in the CC-BY-4.0 license.

%\linenumbers

\section{Introduction}
\label{sec:introduction}
\begin{figure}[hbt!]
    \centering
    \includegraphics[width=0.7\textwidth]{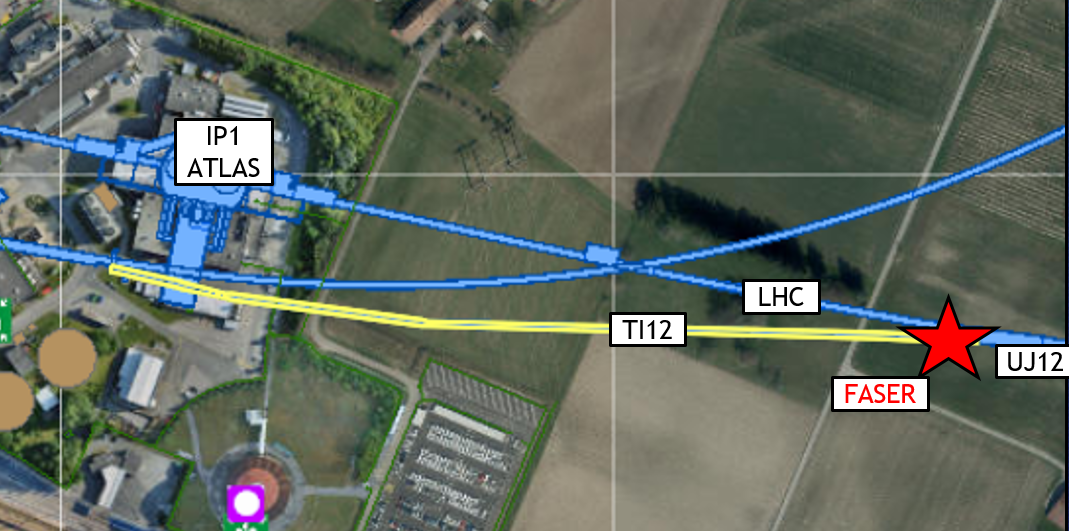}
    \caption{The location of the FASER detector in TI12 tunnel, located 480~m downstream of the ATLAS interaction point along the beam collision axis.}
    \label{fig:TI12}
\end{figure}

FASER (the ForwArd Search ExpeRiment) was proposed in 2017~\cite{Feng:2017uoz} to search for new particles at the LHC at CERN.
The experiment is placed in the unused service tunnel, TI12, situated 480~m downstream from the ATLAS interaction point on the beam collision axis as shown in \Cref{fig:TI12}. 
TI12 was formerly used for connecting the Large Electron-Positron (LEP) collider to the Super Proton Synchrotron (SPS).
Excavation of the floor of TI12 provides a 5.5~m-long and up to 1.5~m-wide trench centred on the beam collision axis, where the FASER detector is installed.
This location enables FASER to search for new weakly-interacting long-lived particles (LLPs) with masses in the MeV to GeV range, produced from the copious inelastic $pp$ scattering events at the LHC.
The detector design has been validated against the benchmark physics process of a dark photon $A'$ decaying into a pair of oppositely charged particles, {\it{e.g}}, $A'\rightarrow e^+e^-$.  
Along with this process, FASER is also sensitive to several other new physics models presented in~\cite{FASER_LLP} which predict LLPs that travel to and decay in the FASER detector into pairs of Standard Model (SM) particles including photons.
In addition to the LLP searches, the FASER location allows to study neutrino interactions of all flavors in an uncharted energy region~\cite{FASERnu}~\footnote{Notably, a 29-kg pilot emulsion detector was temporarily installed in this location during LHC running in 2018, which has been used to observe the first neutrino interaction candidates at a collider~\cite{FistNuInt}.}. 
The experiment location greatly reduces background from the SM entering FASER from the LHC collisions, since such particles would have to traverse the LHC magnets, absorbers and 90~m of rock. 

\begin{figure}[hbt!]
    \centering
    \includegraphics[width=1.0\textwidth]{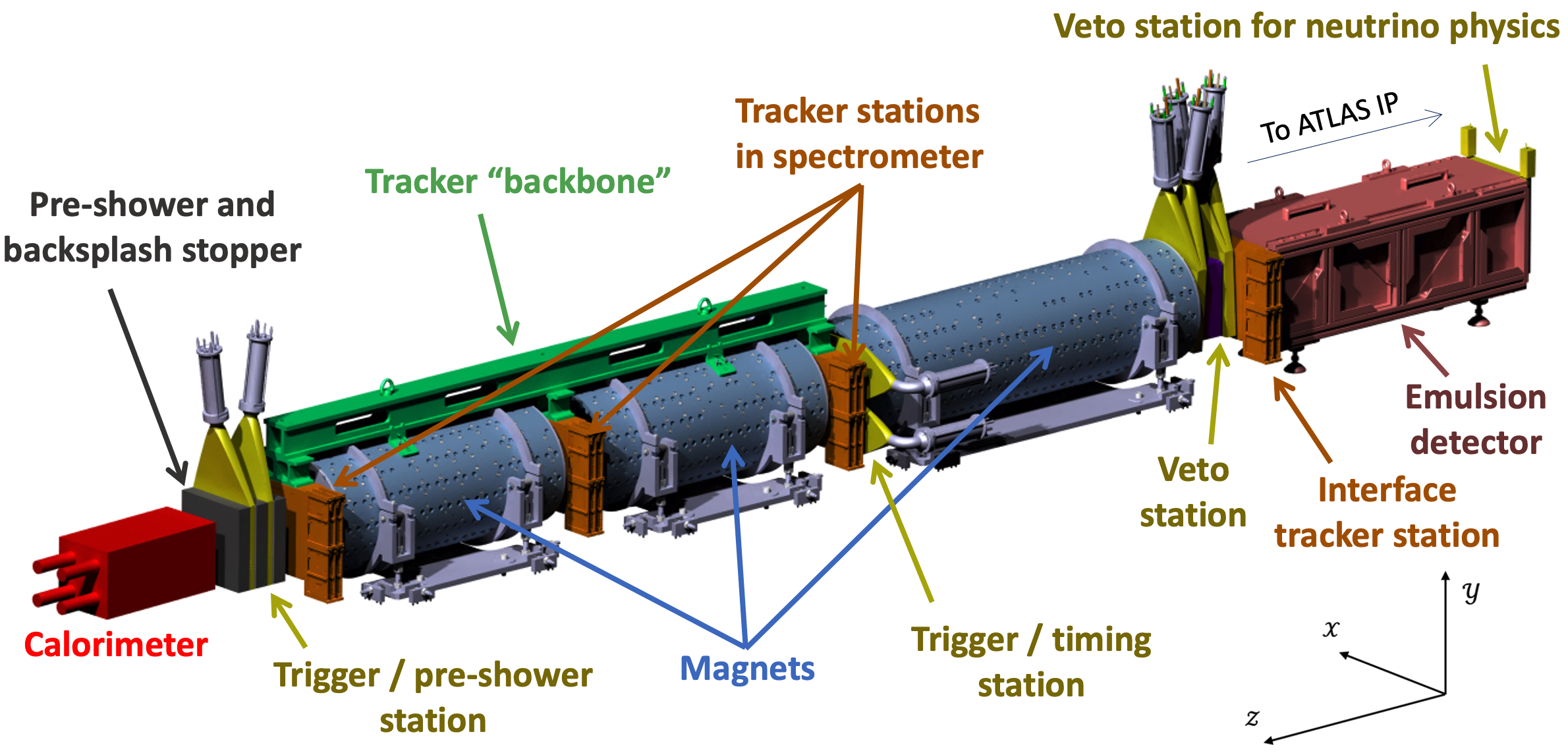}
    \caption{A sketch of the FASER detector. The components labelled {\it Interface tracker station}, {\it Emulsion detector}, and {\it Veto station for neutrino physics} have not yet been installed.}
    \label{fig:FASER}
\end{figure}

A sketch of the FASER detector with the coordinate system is shown in \cref{fig:FASER}.~\footnote{Note, the figure shows some detector components that have not yet been installed.}
The FASER detector for LLP searches was installed in Spring 2021 and is currently being commissioned towards data taking during LHC operation in 2022-24 (Run 3).
As a decay volume for LLP searches, the detector has a 1.5~m-long and 0.55~T dipole magnet with a 10~cm radius aperture to separate the pair of charged particles arising from the LLP decay. Due to the limited space and accessibility to TI12, a permanent magnet with a Halbach array design is used. 
A veto scintillator station and a timing scintillator station are placed upstream and downstream of this to ensure that the LLPs decay inside the decay volume. 
Just behind the timing scintillator station, three tracker stations and two 1~m long magnets are alternately installed to form a tracking spectrometer allowing to measure the position and momentum of charged particles.
These 1~m-long magnets have the same design as the 1.5~m-long decay volume magnet except for the length. 
The tracking stations are situated about 6.5~cm away from the magnet opening and therefore are within the stray field of the permanent magnets with a maximum field in the station of $\sim60$~mT.
Since the LLPs produced in the forward direction of the LHC collisions are light and have energy at the TeV-scale, the pair of charged particles produced in their decay is highly collimated. 
In order to resolve these particles the tracker stations, which each consist of three tracker planes, must have a position resolution of better than $\mathcal{O}$(\SI{100}{\micro\metre}).
For example, for an LLP with mass $m = 100~\textrm{MeV}$ and energy $E = 2~\textrm{TeV}$ decaying inside the decay volume, the separation between the decay products at the first tracker station can be $\mathcal{O}$(\SI{200}{\micro\metre}). 
Although for the FASER LLP searches precise measurement of the signal particle momentum is not needed, the detector is designed to be able to measure charged particle momenta up to a few 100~GeV, with the performance determined by the relative alignment between the tracker stations.
The tracker stations consist of silicon microstrip detectors with a pitch of \SI{80}{\micro\metre}, which fully cover the aperture of the magnets; this paper will focus on details of these tracking detectors.
The most downstream detector is a pre-shower scintillator station and an electromagnetic calorimeter with a depth of 2 and 25 radiation lengths ($X_{0}$), respectively, which will allow to discriminate electrons from muons and to measure the electromagnetic energy.  
A trigger signal for data taking is provided from all the scintillator stations and the calorimeter.
Based on FLUKA~\cite{Ferrari:2005zk,Bohlen:2014buj} simulations and {\textit{in situ}} measurements~\cite{FASERLoI,FASERTP}, a trigger rate of around 500~Hz is expected for a luminosity of $2\times10^{34}$~cm$^{-2}$s$^{-1}$.\footnote{This is the expected peak luminosity during LHC Run 3.} 
This rate is dominated by high-energy muons penetrating through the 90~m of rock to get to FASER. Details of the trigger system are described in~\cite{FASER:2021cpr}. 
All components are integrated and have been commissioned~\footnote{The results of the commissioning tests have been used to determine which modules were installed in which position to maximize the physics performance.}, as described in \cref{sec:commissioning}.

To allow neutrino measurements at FASER, the detector will be augmented with an additional veto scintillator station, a 1.1-ton tungsten/emulsion detector that works as the target for the neutrino interactions, and an interface tracker. These elements will be installed in front of the FASER detector (described above) in late 2021.
The interface tracker will enable tracks from a neutrino interaction in the emulsion detector to be matched to events in the tracker stations. 
This will allow the measurement of both the muon neutrino and muon anti-neutrino interaction cross sections by using the charge of the muon reconstructed in the FASER spectrometer.
The interface tracker will also improve the background rejection and energy reconstruction for the neutrino analysis. 
The interface tracker has an identical design to the tracker stations used in the FASER spectrometer. FASER will therefore have four tracker stations in total, but this paper focuses on the three tracker stations used in the FASER spectrometer and already installed in TI12 in March 2021.

The layout of this paper is as follows. 
\Cref{sec:TrackerStation} presents the mechanical design of the tracker stations including details of the silicon strip detectors and on-detector electronics. The power supply, cooling, and data acquisition systems are outlined in \cref{sec:equipments}.
The interlock and detector control systems are summarized in \cref{sec:DCS}. 
All components are integrated and have been commissioned~\footnote{The results of the commissioning tests have been used to determine which modules were installed in which position in the detector to maximize the physics performance.}, as described in~\cref{sec:commissioning}. Finally, conclusions and outlook are given in \cref{sec:conclusion}.

\section{Tracker stations}
\label{sec:TrackerStation}

The FASER tracker consists of 72 double-sided silicon microstrip modules arranged in three stations, each station being composed of three planes with eight modules per plane. Given the short timescale to build the entire tracker, spare barrel modules of the ATLAS Semiconductor Tracker (SCT) have been used. The SCT is one of the three sub-systems composing the Inner Detector~\cite{ATLAS:2010ylv}, the silicon tracker of the ATLAS experiment at the LHC~\cite{ATLAS:2008xda}. 

%==============================================
\subsection{Silicon microstrip modules}
\label{sec:modules}
%==============================================
The SCT barrel module~\cite{Abdesselam:2006wt} consists of four identical single-sided silicon microstrip sensors glued in pairs on the two sides of a central baseboard. A copper/polyimide flex hybrid carrying the readout electronics is glued on top of one of the sensors. Each sensor has $p^+$ strip  implants on a high-resistivity ($>\SI{4}{\kilo\ohm\centi\metre}$) \SI{285}{\micro\metre}-thick $n$-type substrate. The strip implants are AC-coupled to aluminum readout strips via a silicon dioxide layer. The High-Voltage (HV) is applied via a $n$-type metalized implant covering the back-plane and the strip implants are biased through polysilicon resistors. Each sensor has a rectangular nominal geometry of $64\times 63.6~\si{\milli\metre\squared}$, with 768 readout strips at a constant pitch of \SI{80}{\micro\metre}. A guard-ring structure surrounding the sensor active area is used to prevent electrical breakdown at large bias voltages. The readout of 128 strip channels is done by means of the Atlas Binary ABCD3TA ASIC~\cite{Campabadal:2005rj}, which comprises the front-end (FE) electronics, a pipeline, buffer, readout and control logic. For each channel, the analog FE consists of a preamplifier, shaper and discriminator. The signal delivered by the preamplifier-shaper circuit has a peaking time of \SI{25}{\nano\second}, enough to ensure a discriminator timewalk of less than \SI{16}{\nano\second} and a double-pulse resolution below \SI{50}{\nano\second} as required for ATLAS operations. Since the discriminator threshold remains common to all channels within the chip, the effective threshold of a given channel corresponds to the offset of its discriminator with respect to the common threshold. A per-channel threshold correction 4-bit DAC allows to compensate for the spread of these threshold offsets. The pipeline, clocked at 40 MHz rate and with a depth of 132 cells, buffers the binary data of all channels for \SI{3.3}{\micro\second}.  Upon reception of a trigger the data from the last three cells are transferred to a second-level derandomizing buffer where data compression is finally applied.

\begin{figure}[t]
\centering
\includegraphics[width=0.75\textwidth]{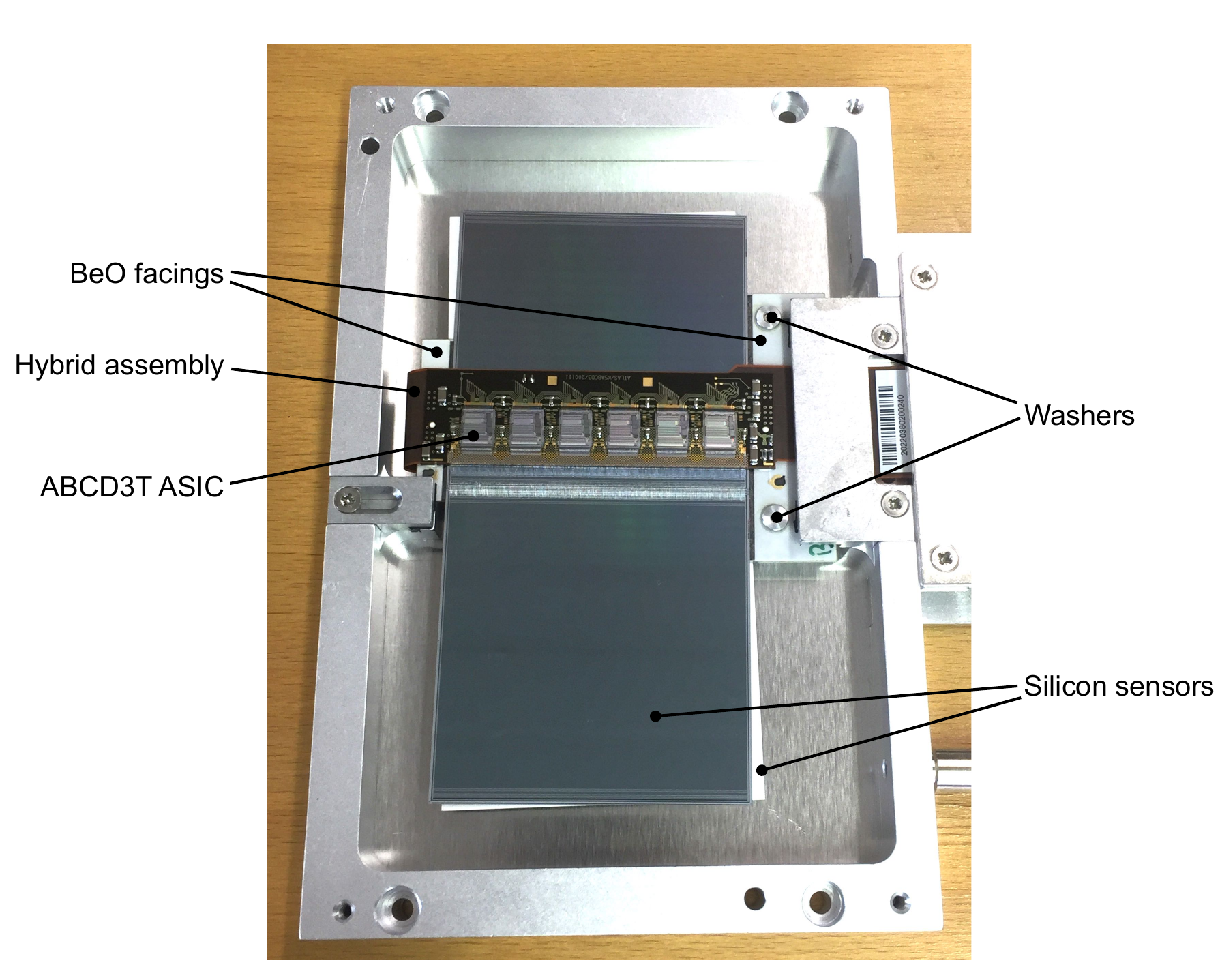}
\caption{Photograph of a SCT barrel strip module.}
\label{f:SCT-photo}
\end{figure}

\Cref{f:SCT-photo} shows a SCT barrel module held inside an aluminum test-box. On each side of the module the sensors are bonded edge-to-edge to create $\SI{\sim12.8}{\centi\metre}$-long readout strips. The total number of readout channels in the module is 1536 (768 per side). A \SI{40}{\milli\radian} stereo angle between the front and back pairs of sensors allows the module to provide a spatial resolution of $\SI{\sim 17}{\micro\metre}$ in the precision coordinate (perpendicular to the strips) and $\SI{\sim 580}{\micro\metre}$ in the non-precision coordinate (parallel to the strips)~\cite{Abdesselam:2006wt}. The flex hybrid, equipped with six ABCD3TA chips per side, is bridged over the sensors via a carbon-carbon substrate. Two NTC thermistors (one per hybrid side) allow to monitor the module temperature. The baseboard, made of Thermal Pyrolytic Graphite (TPG) with an excellent in-plane thermal conductivity and low radiation length, provides the mechanical support to the sensors and allows to dissipate the heat generated by the FE electronics. The hybrid is attached to beryllia (BeO) facing plates located on the two ends of the TPG baseboard.

The modules used for the FASER tracker have been selected among the existing spares of the SCT barrel module production that was completed in 2004. Since then, the modules were stored at CERN in individual sealed bags within a controlled environment. Some of these spare modules failed the initial quality assurance by the SCT collaboration, typically because there were more than 1\% defective channels, high leakage current (above \SI{4}{\micro\ampere}) or early breakdown before \SI{500}{\volt} (the maximum voltage expected after 10 years of operation in ATLAS). Given the much lower radiation levels expected in FASER compared to ATLAS\footnote{A total ionizing dose less than $5\times 10^{-3}$ Gy per year and a total fluence less than $5 \times 10^7$ \mbox{1-MeV-neutrons} equivalent $\rm{n_{eq}/cm^2}$ per year are estimated from simulations with FLUKA and confirmed with {\it{in-situ}} measurements in the TI12 tunnel~\cite{FASERTP}. The SCT barrel module is designed to maintain performance even after a total ionizing dose of 100 kGy and a total fluence of $2 \times 10^{14}$ \mbox{1-MeV-neutrons} equivalent $\rm{n_{eq}/cm^2}$~\cite{Abdesselam:2006wt}.}, the nominal operating voltage for the modules is set to \SI{150}{\volt} for the entire detector lifetime. This bias voltage is large enough to ensure full depletion of the sensors, and it is not expected to be increased during the lifetime of the experiment due to the lack of cumulated radiation damage. Therefore, for FASER, out of modules operational up to 300~V, the ones with the lowest number of defects have been chosen. The IV curves have been measured at different stages of the tracker assembly (more details are given in \cref{sec:commissioning}).

%==============================================
\subsection{Tracker plane} 
\label{sec:tracker-plane}
%==============================================

\begin{figure}[th]
\centering
\includegraphics[width=0.65\textwidth]{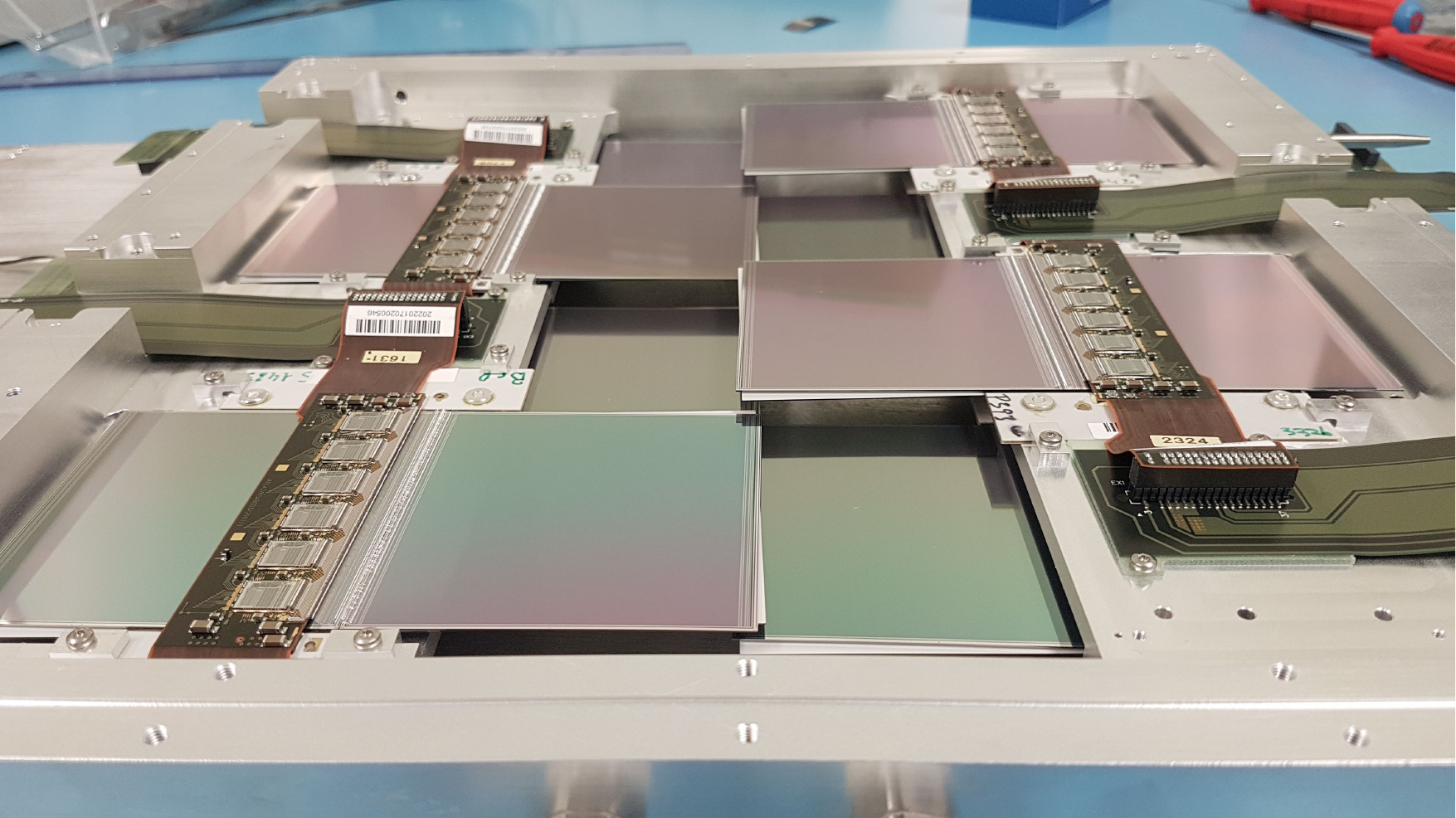}
\caption{Photograph of a tracker plane with all eight SCT modules installed. The beam axis is perpendicular to the plane.}
\label{f:plane-photo}
\end{figure}

\begin{figure}[th]
\centering
\includegraphics[width=0.52\textwidth]{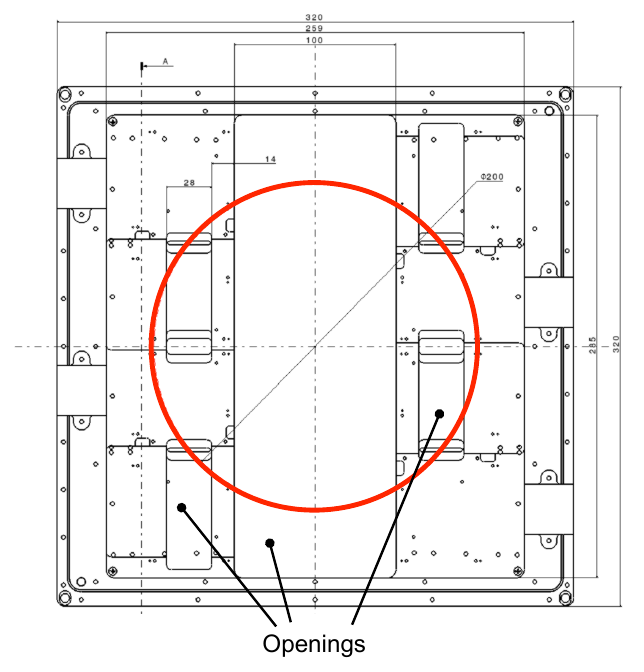}
\caption{CAD view of a frame. The openings of the frame are intended to minimize materials inside the \SI{200}{\milli\metre}-diameter magnet aperture illustrated by the circle.}
\label{f:frame-cad}
\end{figure}

A tracker plane consists of an aluminum frame (AW-5083) holding eight SCT barrel modules as shown in \cref{f:plane-photo}. 
The modules are arranged by four on each side (front and back) of the frame to minimize non-sensitive material in the central region. 
To measure the momentum of the charged particles separated by the magnetic field, the modules are oriented with the strips perpendicular to the $y$-axis (sensitive coordinate).
The distance between modules (closest sensors along the out-of-plane direction) is \SI{2.4}{\milli\metre}, and the active area overlap (in-plane, along the strip-length) is \SI{2}{\milli\metre}. 
The overall active area in the tracker plane is $\SI{240}{\milli\metre}\times \SI{240}{\milli\metre}$, which covers the $\SI{200}{\milli\metre}$-diameter magnet aperture as illustrated in \cref{f:frame-cad}. The frame has a size of $\SI{320}{\milli\metre}\times \SI{320}{\milli\metre}\times \SI{31.5}{\milli\metre}$. In order to minimize the material in front of the silicon detectors, the frame is cut-out for most of the active area within the acceptance of the magnet aperture. More details about the material distribution in the active area of a tracker station are given in \cref{sec:tracker-stations}.

\begin{figure}[thbp]
\centering
\subfloat[\label{f:plane-bare}]{\includegraphics[width=0.46\textwidth]{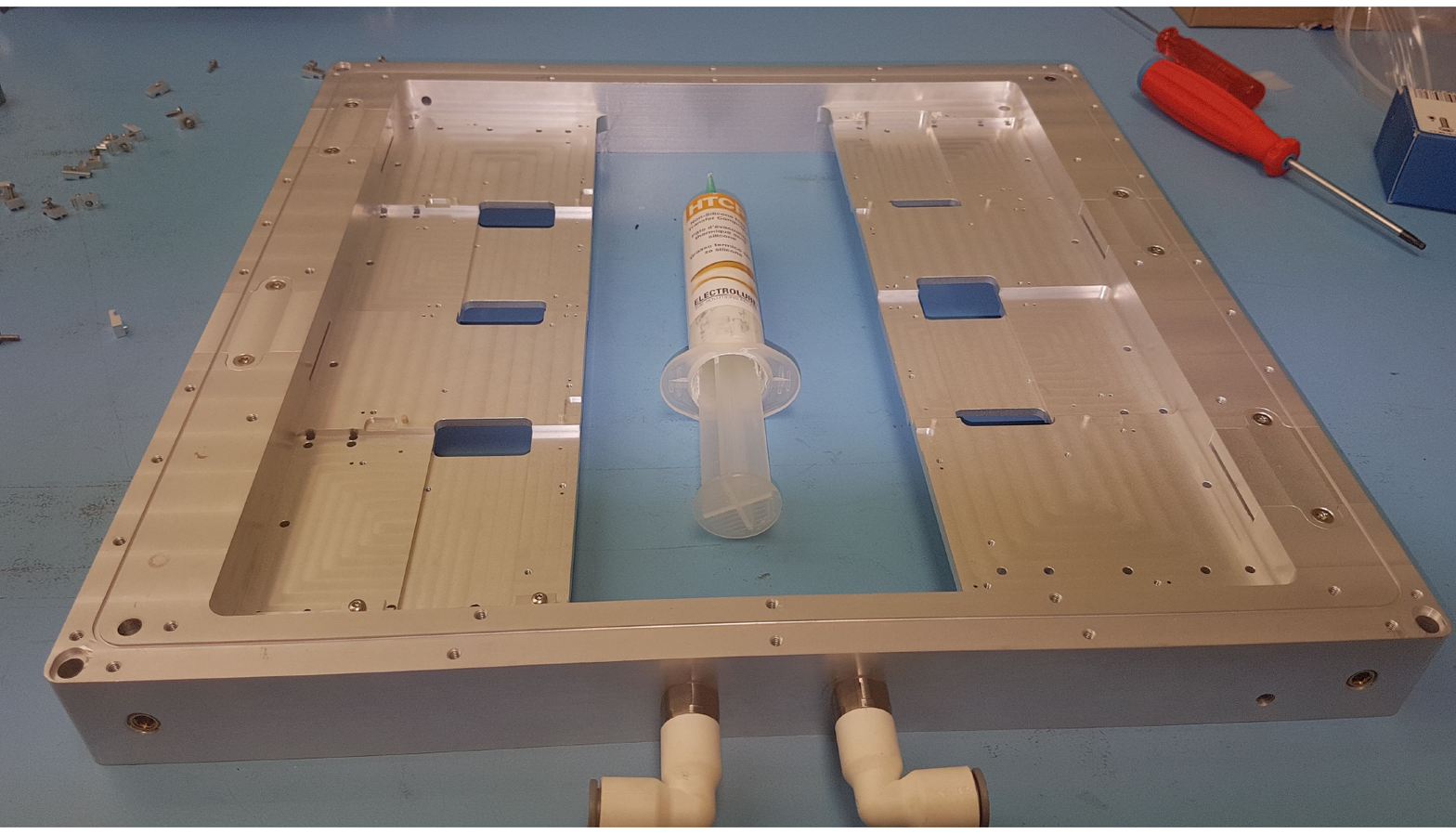}}
\hfil
\subfloat[\label{f:plane-module-fixation}]{\includegraphics[width=0.48\textwidth]{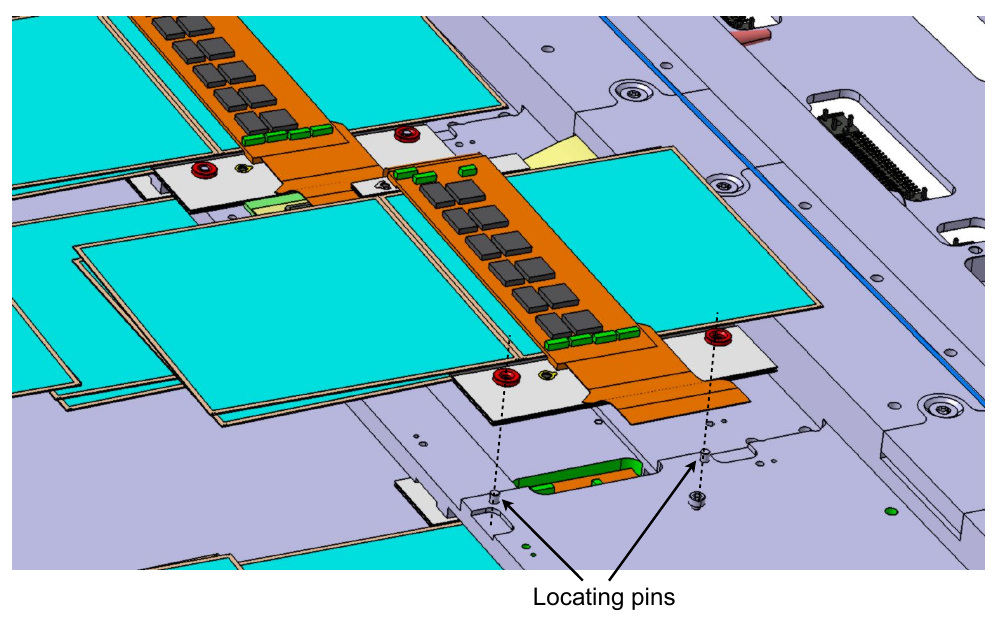}}\\
\subfloat[\label{f:plane-module-clamps}]{\includegraphics[width=0.45\textwidth]{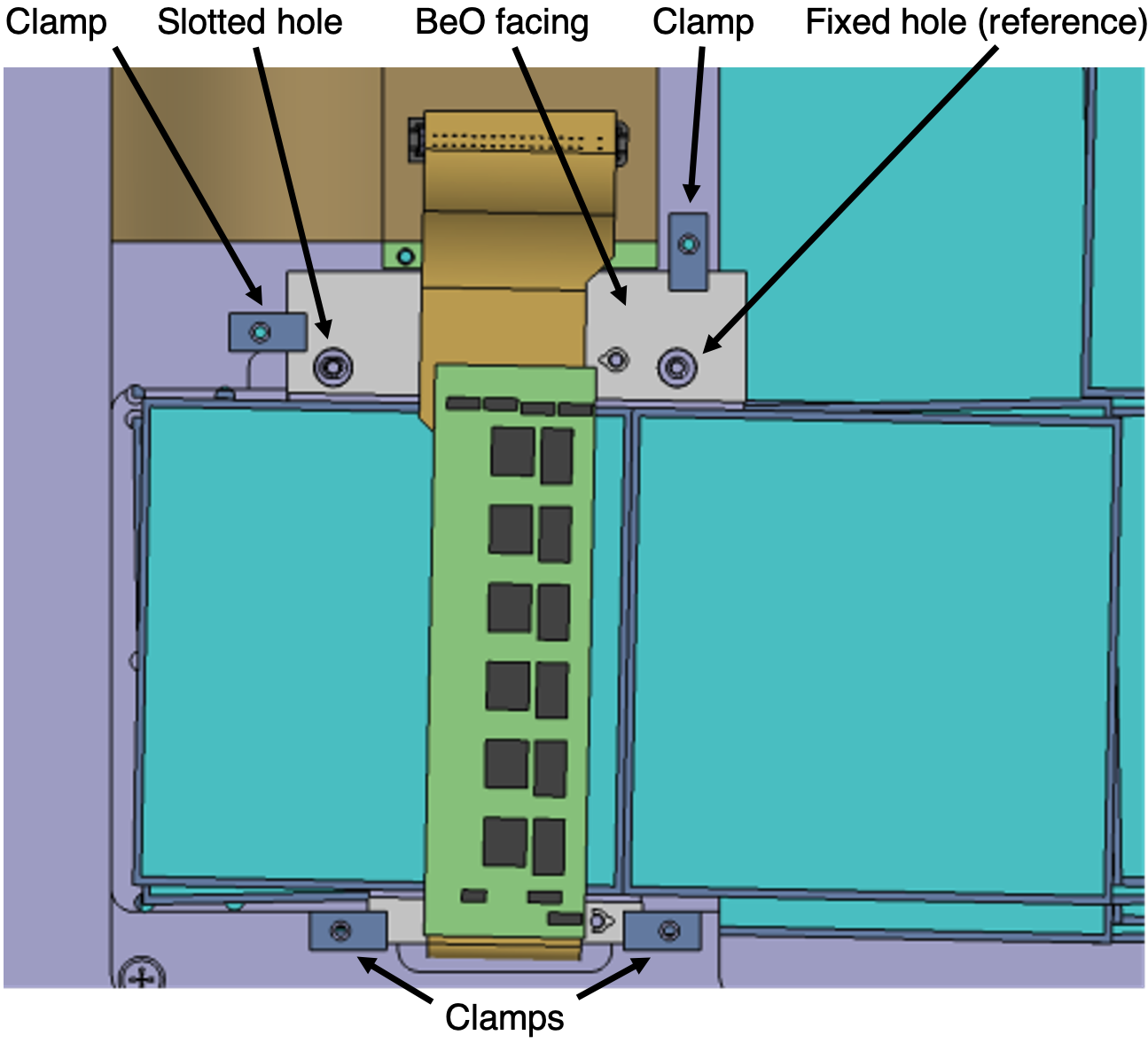}}%
\hfil
\subfloat[\label{f:plane-cooling-channel}]{\includegraphics[width=0.5\textwidth]{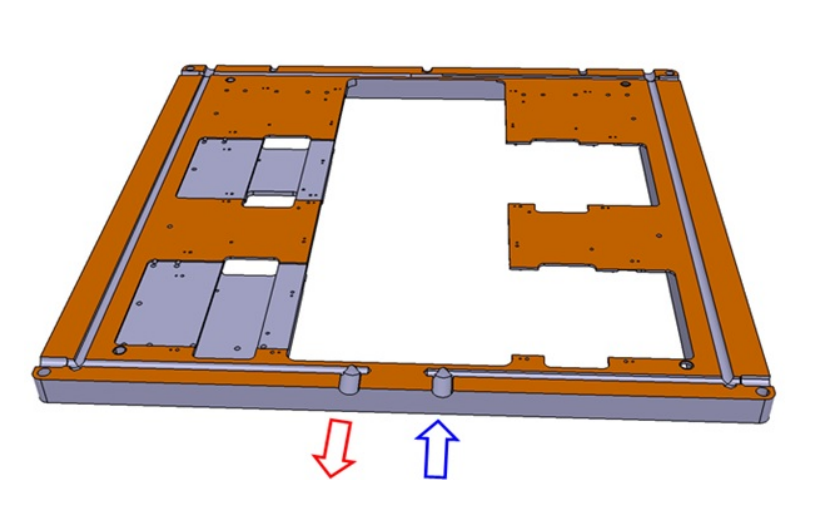}}
\caption{(a) Bare aluminum frame, (b) exploded CAD view of a SCT module while being positioned, (c) exploded CAD view of a module after its final fixation onto the frame using clamps and (d) mid-plane cross section showing the inner cooling channel. In (d), the arrows indicate the direction of the water flow.}
\label{f:plane}   
\end{figure}

The aluminium frames are CNC (Computer Numerical Control) machined (\cref{f:plane-bare}) and then individually surveyed to check the quality and precision of the production process. The SCT modules are positioned in the frame by using two \SI{1.8}{\milli\metre}-diameter PEEK 1000 pins (\cref{f:plane-module-fixation}). One locating pin provides the global precise positioning while a second pin (to be inserted into the slotted washer of the SCT module) is used for the module orientation. The PEEK material allows for a smooth and easy handling of the modules during insertion-removal operations. The modules are then pressed down to the frame by means of four small aluminum clamps fixed by four stainless steel M2 screws (\cref{f:plane-module-clamps}). A \SI{5}{\milli\metre}-diameter inner cooling channel, integrated by design into the frame (\cref{f:plane-cooling-channel}), provides the thermal path to extract by direct conduction through the water flow the heat generated by the FE electronics. 
A heat conducting thermal paste (Electrolube HTCP-20S) is applied at the contact surface between the frame and the bottom side of the BeO facings to improve the thermal contact between the modules and the aluminum frame.
The coolant temperature is set at \ang{15}C.
Absolute humidity for condensation at \ang{15}C is 12.8\%.
The frame also contains an inlet for dry air to be flushed, which would achieve the humidity inside at around 1-2\%.

\begin{figure}[th]
\centering
\includegraphics[width=0.65\textwidth]{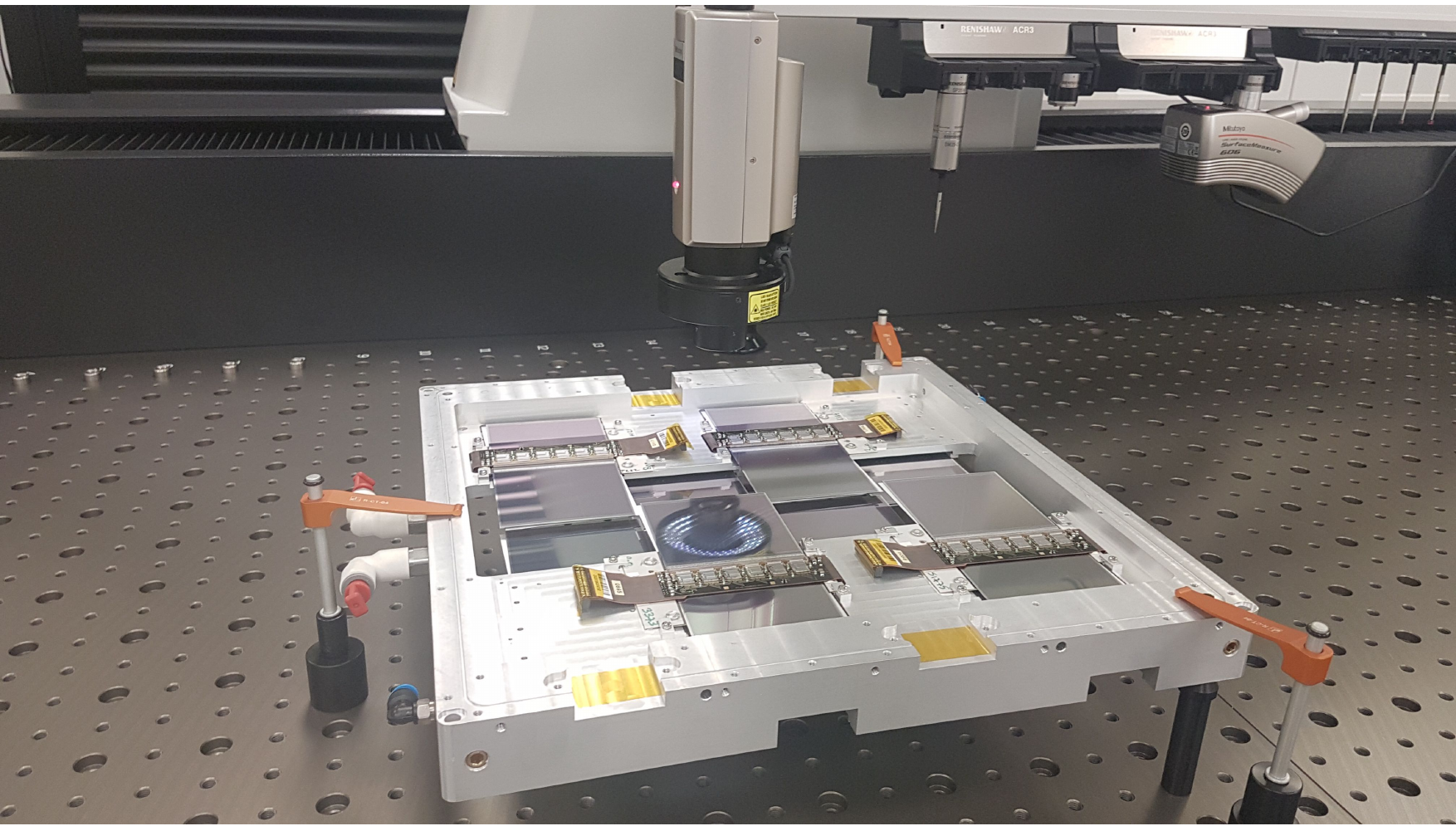}
\caption{Metrology of a fully assembled tracker plane.}
\label{f:metrology}
\end{figure}

The metrology of each plane combines measurements performed with a mechanical touch-probe and an optical camera. Each frame is equipped with four stainless steel targets (one in each corner) used to define the plane reference coordinate system. The targets are visible from both sides, allowing to correlate measurements done on each side of the plane separately. The silicon sensors of the SCT modules include a set of fiducial marks used for the mechanical alignment during module assembly. Some of the fiducial marks are measured in 3D with respect to the plane reference system. All assembled planes were measured at the University of Geneva using a Mitutoyo CRYSTA-Apex S CNC coordinate measuring machine with automatic probe changer (\cref{f:metrology}). The precision of the metrology machine for in-plane and out-of-plane measurements is \SI{5}{\micro\metre} and 10-15~\si{\micro\metre}, respectively. It is found that all frames are within the required tolerances ($\pm \SI{20}{\micro\metre}$) with respect to the CAD manufacturing drawings. A maximum deviation of \SI{100}{\micro\metre} is found for the positioning of the SCT modules with respect to the CAD model (corresponding to a perfect alignment). 
This maximum deviation accounts for the combined effect of machining tolerances, SCT modules positioning errors and assembly precisions. 
The required precision of the alignment is $\mathcal{O}$(\SI{50}{\micro\metre}). The initial alignment will be based on the metrology data, but can be improved and the stability monitored using track data during FASER operations.

%==============================================
%\subsection{On-detector electronics} 
\subsection{Pigtail and patch-panel} 
\label{sec:on-detector-electronics}
%==============================================

\begin{figure}[tb]
\centering
\subfloat[\label{f:pigtail}]{\includegraphics[width=6cm]{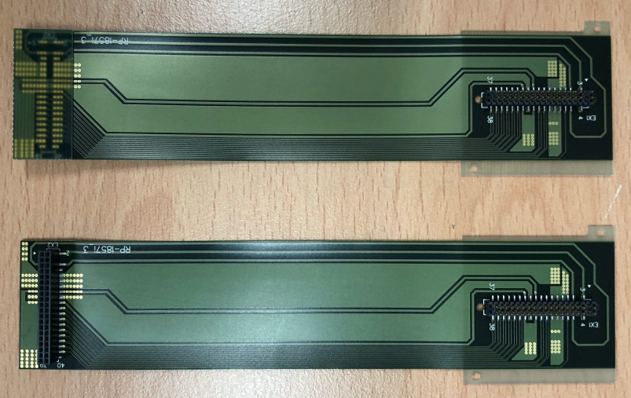}}\\
\subfloat[\label{f:patch-panel}]{\includegraphics[width=12cm]{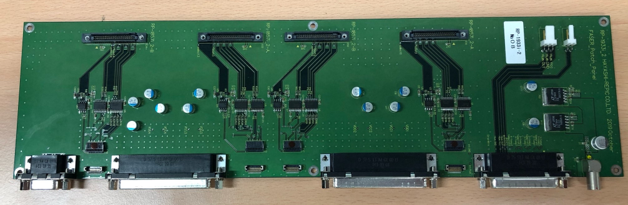}}
\caption{(a) The two types of pigtails with connectors to the patch-panel at opposite sides and (b) the patch-panel.}
\label{fig:pigtail-pp}
\end{figure}

\begin{figure}[tb]
\centering
\includegraphics[width=12cm]{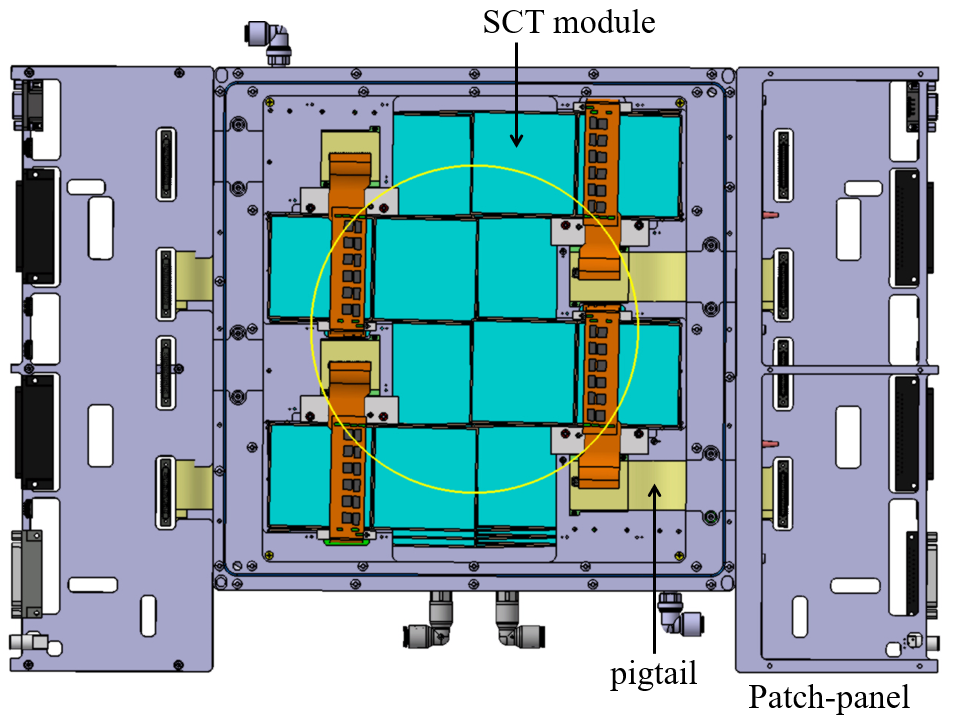}
\caption{Schematic view of the tracker plane. One pigtail is connected to each hybrid on the SCT module. Note that adjacent SCT modules are mounted on opposite sides of the FASER module frame. The four pigtails in on each side are connected to one patch-panel.}
\label{fig:FPCB-PP}
\end{figure}

A flexible printed circuit board (called the "pigtail") has been developed to route the electric lines of the SCT module to the outside (\cref{f:pigtail}). Four pigtails (one per module) at each side of the tracker plane are connected to a single patch-panel (\cref{f:patch-panel}).
The connectors on the patch-panel are mounted on the same side while the side of the module is arranged alternatively in the tracker plane as shown in \cref{fig:FPCB-PP}. For that reason, two types of the pigtail were developed.
The size of the pigtail is 6~cm~$\times$~15~cm with \SI{150}{\micro\metre} thickness. The connector\footnote{SFMC-120-L3-S-D manufactured by SAMTEC, Inc} on the module side is oriented with 90 degrees on the pigtail to connect with the hybrids on the module as shown in \cref{fig:FPCB-PP}. 

The tracker is electrically connected to the powering and readout systems (see \cref{sec:equipments}). The patch-panel (\cref{f:patch-panel}) was developed as the interface between them. A patch-panel is placed at each side of the tracker plane and fixed on the aluminum frame. The size of the patch-panel is 10~cm~$\times$~30~cm with 2~cm thickness including the height of connectors.
The patch-panel is designed to be thinner than the thickness of the frame, so that the tracker planes can be stacked to assemble the tracker station (see \cref{sec:tracker-stations}).
The Tracker Readout Board (TRB, see \cref{sec:DAQ}) in the readout system is connected to two patch-panels of the same plane via eight twinax Firefly cables\footnote{ECUE-08-300-T2-FF-01-1 manufactured by SAMTEC, Inc}, {\em i.e.} one cable per SCT module. The cable provides all LVDS lines necessary for operation and readout of the ABCD3TA chips on the module. LVDS repeaters are mounted for all LVDS lines on the patch-panel to transmit signals along 3~m of the cable. In addition, LVDS receivers are placed to receive the operation signals for the ABCD3TA chips from the TRB. HV and Low-Voltage (LV) as well as their return lines are provided from a HV splitter board and a LV protection board to bias the sensors and power the ABCD3TA chips, respectively (see \cref{sec:powering}). Every patch-panel is connected with the HV splitter board, biasing the sensors on the four modules with one HV channel. LV for the analog and digital circuits in the ABCD3TA chips is provided to each module separately. The patch-panel works as the interface for the NTCs on the modules and temperature and humidity sensors on the plane frame with the Tracker Interlock and Monitoring Board (TIM, see \cref{sec:TIM}) which is used for the environmental monitoring. In addition, there are 5~V power lines on the patch-panel for the LVDS repeaters and receivers as well as a line to transfer an interlock signal generated by the TIM to the LV power supply.

%==============================================
\subsection{Tracker stations}
\label{sec:tracker-stations}
%==============================================

\begin{figure}[tb]
\centering
\subfloat[\label{f:station-exploded}]{\includegraphics[width=0.48\textwidth]{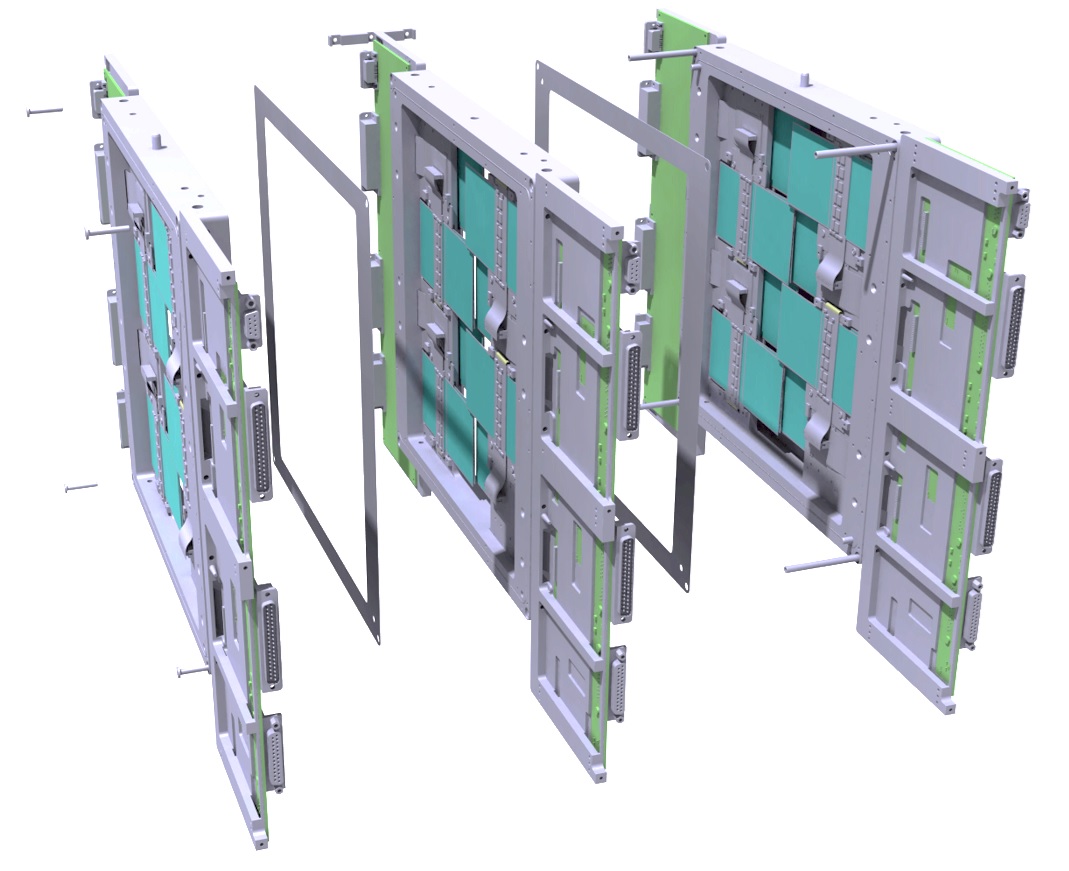}}
\hfil
\subfloat[\label{f:station-photo}]{\includegraphics[width=0.48\textwidth]{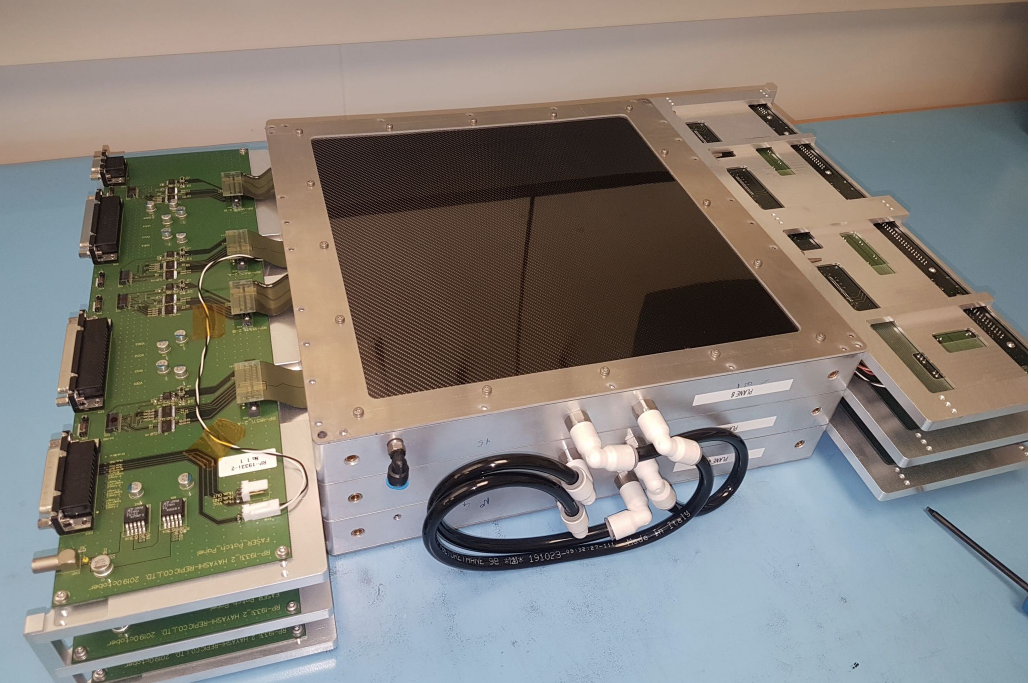}}%
\caption{(a) Exploded CAD view of a tracker station and (b) photograph of a fully assembled station. The black cover on top of the station is a carbon-fibre plate. The cooling loops of the three planes are connected together in series, so that per station there is only one inlet and one outlet for the cooling fluid. In this photograph, the inlet and outlet are also connected by a temporary pipe.}
\label{f:station}   
\end{figure}

A tracker station is an assembly of three planes (including the patch-panels) as shown in \cref{f:station}. 
Each station is flushed with dry air to reduce the relative humidity ensuring safe operating conditions against dew point as discussed in \cref{sec:tracker-plane}.
One inlet for the dry air is used for the whole station, while the other two are sealed by screws.
%The different patch-panels are mounted and stay within the \SI{32}{\milli\metre} total thickness of each plane. 
All aluminium parts are post-treated with a trivalent chromium passivation\footnote{SURTEC-650 coating} to prevent any corrosion by the corona effect that may occur after putting the frames in contact during the station assembly. In addition, an O-ring sealing joint between frames provides a good tightness to keep the humidity inside the station as low as possible (typically about 1-2\%). The station is assembled from three planes only after the metrology and full commissioning of each individual plane is completed. The inter-plane alignment is done via two high-precision pins located in \SI{5}{\milli\metre}-diameter H7 holes while the fixation between two adjacent planes is done via four M5 screws. The two end covers that close the station volume are made of $\SI{400}{\micro\meter}$-thick carbon-fibre plates (standard T300 fibers). Without cables, the total weight of one station is about \SI{15}{\kilogram}.

\begin{figure}[t]
\centering
\subfloat[\label{f:simu-plane}]{\includegraphics[width=0.48\textwidth]{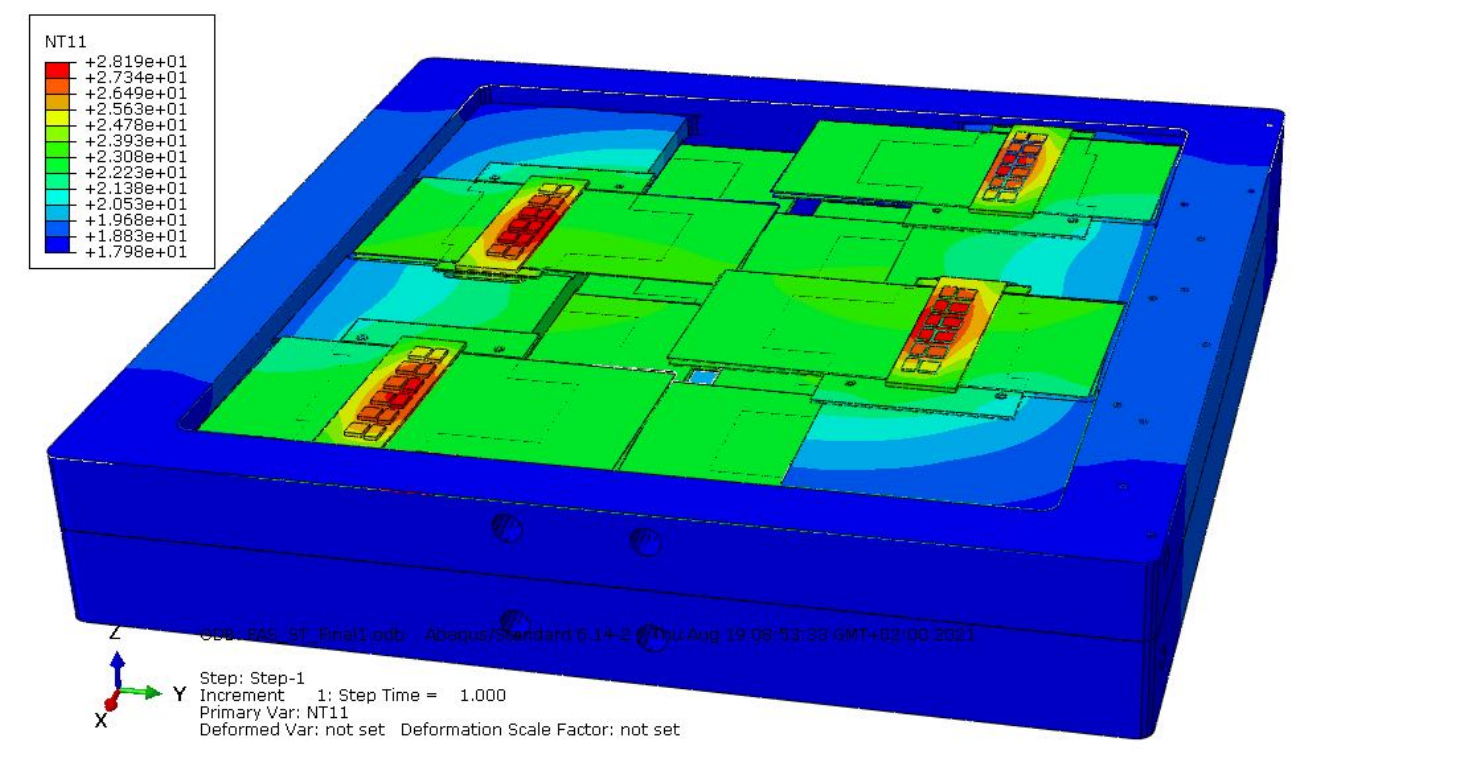}}
\hfil
\subfloat[\label{f:simu-module}]{\includegraphics[width=0.48\textwidth]{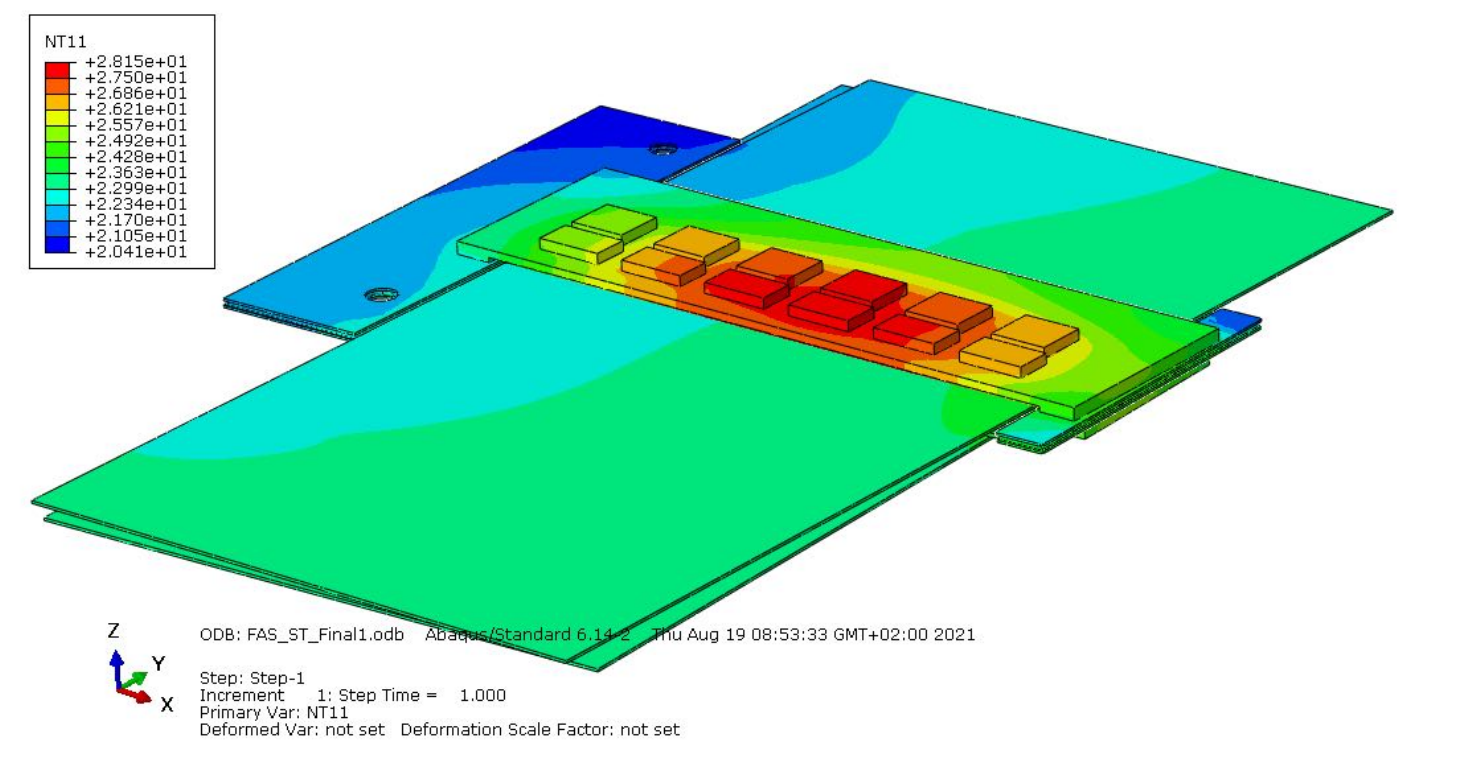}}%
\caption{FEA simulations on a complete station. (a) two planes in the station and (b) close up view on a single module. The maximum temperature estimated on the module is \SI{28}{\degreeCelsius} (on the readout electronics). See text for additional details on the different simulation parameters.}
\label{f:fea-simus}   
\end{figure}

Various Finite Element Analyses (FEA) simulations have been carried out to check the thermal performance of the system throughout the evolution of the frame design.
Over-heating of the SCT modules can cause problems in the mechanical integrity and alignment of the SCT modules due to the glass-transition of glues.
The glue\footnote{ARALDITE 2011 manufactured by Huntsman Corporation} used for the SCT module assembly should have the lowest glass-transition temperature around \SI{35}{\degreeCelsius}, which should be the absolute maximum temperature for the SCT modules. The safety scheme to ensure this is described in \cref{sec:DCS}.
\Cref{f:fea-simus} shows the FEA of two planes in a station. The different simulation parameters have been set to match the testing conditions during the plane and station commissioning at CERN. In particular, the water temperature for cooling the tracker station has been fixed at \SI{15}{\degreeCelsius}, the water flow at \SI{3}{\litre\per\minute} (for a heat transfer coefficient of water of $\SI{500}{\watt\per\metre\squared}$), and the outside air convection at \SI{23}{\degreeCelsius}. The FEA gives a maximum temperature on the FE chips of $\sim\SI{28}{\degreeCelsius}$, neglecting the temperature rise within the water channel due to the heat load (estimated to be +\SI{0.6}{\degreeCelsius} for \SI{3}{\litre\per\minute}). The FEA results are in good agreement (within 2-\SI{3}{\degreeCelsius}) with the measurements taken during the commissioning at CERN. Given the good modeling of data measurements by the FEA, the temperatures on the silicon sensors are estimated to be in a comfortable range between 21 and \SI{23}{\degreeCelsius}.

\begin{figure}[ht]
\centering
\includegraphics[width=0.9\textwidth]{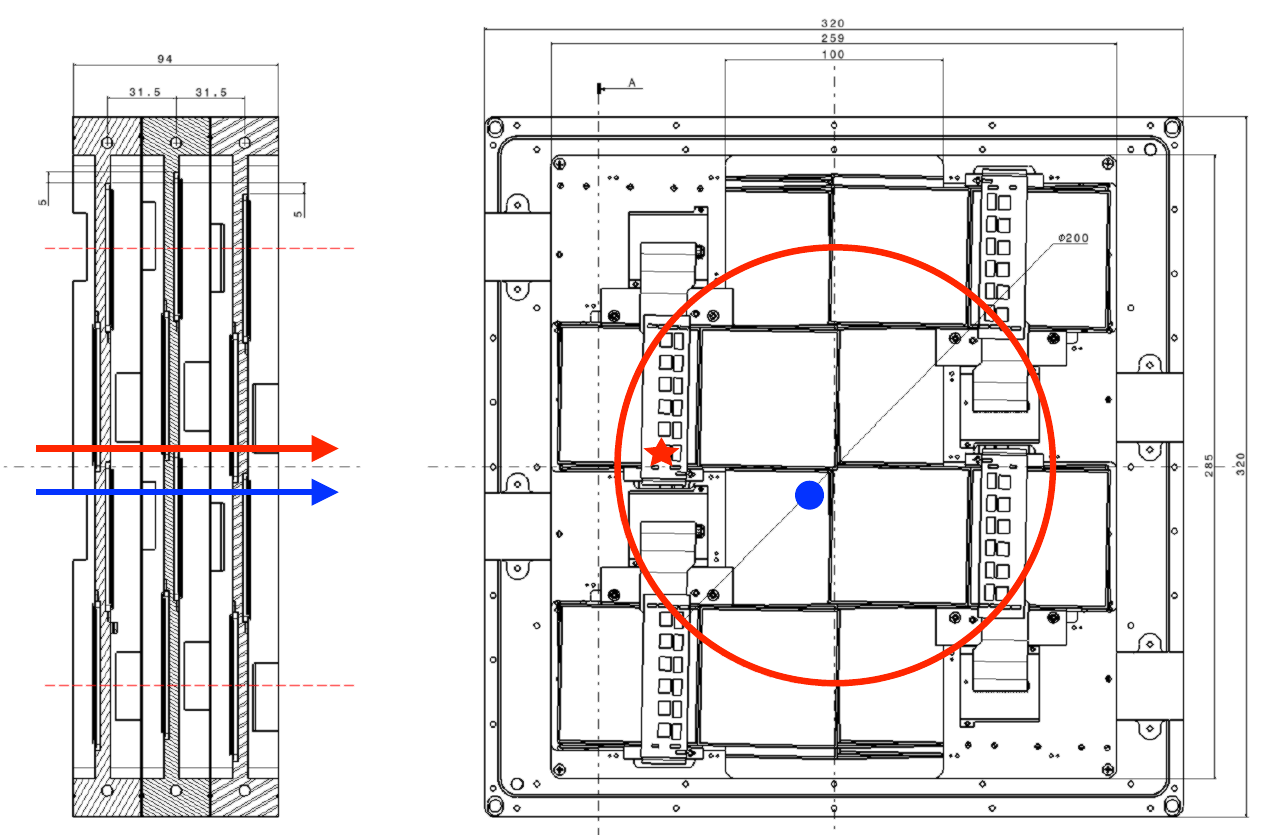}
\caption{Side (left) and front (right) views of a tracker station. The circle represents the \SI{200}{\milli\metre}-diameter magnet aperture, {\it{i.e.} the acceptance of the tracker station}. The blue dot and red star markers respectively correspond to the reference positions for material calculation in the central and edge regions.}
\label{f:station-material}
\end{figure}

\Cref{f:station-material} shows the side and front CAD views of a station. 
Within the station the three planes are staggered along the $y$-axis, with a relative shift of the middle (last) plane of $+\SI{5}{\milli\metre}$ ($-\SI{5}{\milli\metre}$) with respect to the first plane.
The staggering of the planes can be seen in the side view.
This configuration ensures at least two 3D reconstructed space-points for a track crossing the station.
Two different reference positions have been considered to estimate the material distribution along the $z$-axis: a {\em central region}, close to the geometrical center of the plane, and an {\em edge region}. \Cref{tab:TrackerMaterial} summarizes the material budget in each case. The central region corresponds to a particle traversing the least amount of material inside the station, {\em i.e.} six silicon sensors and two carbon-fibre covers that account for a total of 2.1\% of a radiation length ($X_{0}$). 
The edge region is a worst-case position that corresponds to a particle traversing the six SCT modules (including sensors, TPG baseboard, flex hybrid with carbon-carbon bridge and readout ASICs), aluminum frames and station covers, accounting for a total of 21.5\%~$X_{0}$. For a benchmark dark photon model (m$_{A'}$=100 MeV, $\epsilon$ = 10$^{-5}$) for dark photons that decay in the FASER magnet aperture, 70\% will be in the low material central region of the tracker. Given the range of particle momentum expected in the experiment the contribution of multiple scattering from the traversed material is expected to be negligible. 

\begin{table}[thb]
    \centering
    \begin{tabular}{|l|c|c|c|c|}
  \hline
Component & Material & Number & \multicolumn{2}{c|}{$X_0$ (\%)}  \\
 & & / station & Central region & Edge region \\
\hline
Silicon sensor & Si  & 6 & 1.8\% & 1.8\%  \\
\hline
 Station Covers & CFRP  & 2 & 0.3\%  & 0.3\%   \\
\hline
SCT module support & TPG  & 3 & -  & 0.6\%  \\
\hline
C-C Hybrid & C (based) & 3 & - & 2.2\%  \\
\hline
ABCD3TA chips & Si & 3 & - & 6.5\%   \\
\hline
Layer frame & Al & 3 & - & 10.1\%   \\
\hline
\bf{Total / station} & \bf{-} & \bf{-} & \bf{2.1\%} & \bf{21.5\%} \\
\hline
    \end{tabular}
    \caption{Amount of material in $X_0$ in the active area of a tracker station for two regions: i) the central region with only the silicon sensor material and ii) the edge region. Details of the material in the SCT module are given in Table 8 of Ref.~\cite{Abdesselam:2006wt}. The numbers are calculated directly from the CAD description of the tracker station.}
    \label{tab:TrackerMaterial}
\end{table}

%==============================================
\subsection{Tracker backbone}
\label{sec:tracker-backbone}
%==============================================

\begin{figure}[t]
\centering
\subfloat[\label{f:backbone-CAD}]{\includegraphics[width=0.5\textwidth]{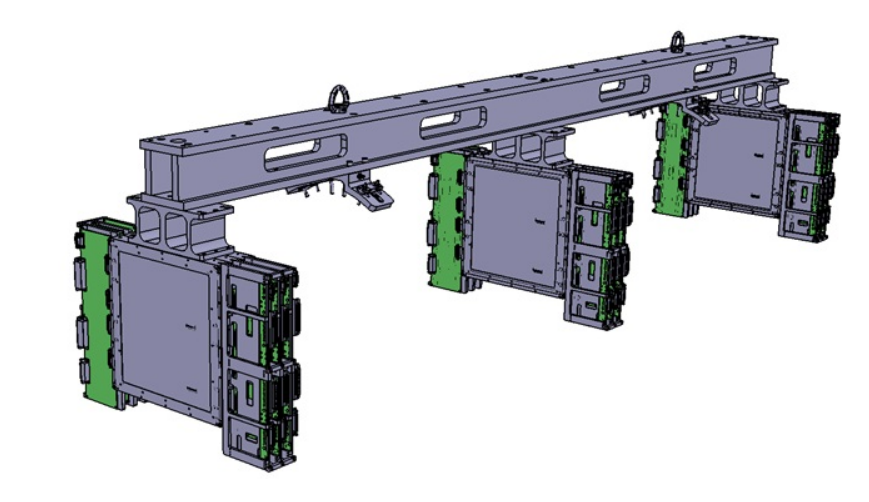}}
\hfil
\subfloat[\label{f:station-interface}]{\includegraphics[width=0.3\textwidth]{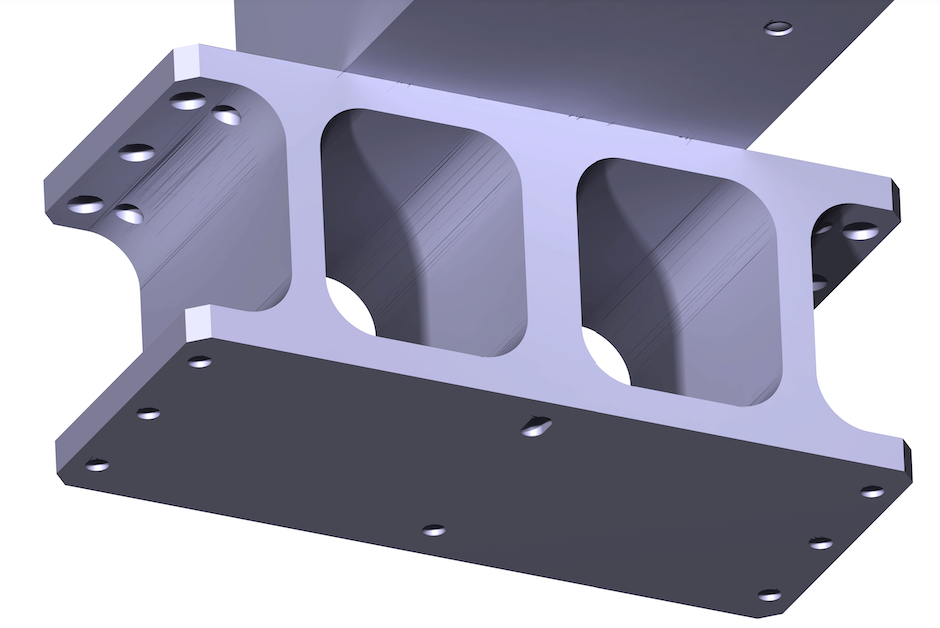}}
\caption{(a) CAD model of the backbone with three stations attached and (b) detail of the station interface.}
\label{f:backbone}   
\end{figure}

The three tracker stations are mounted into the FASER detector with an aluminum structure (AW-5083) called the {\em backbone} as shown in \cref{f:backbone-CAD}. Each station is fixed to the backbone via an aluminum station interface (\cref{f:station-interface}) with six M6 screws and two locating pins. The backbone has been CNC machined in one go for what concerns its key parts (stations interface areas, pin holes) to optimize the final precision. It has been assembled at the University of Geneva and surveyed by the CERN group to check its overall flatness, parallelism and straightness. The survey results show a \SI{0.13}{\milli\metre} flatness over the three station interfaces, each one being \SI{0.04}{\milli\metre} without load and \SI{0.10}{\milli\metre} when loaded with the station weight. 

\begin{figure}[htbp]
\centering
\subfloat[\label{f:backbone-transport}]{\includegraphics[width=0.55\textwidth]{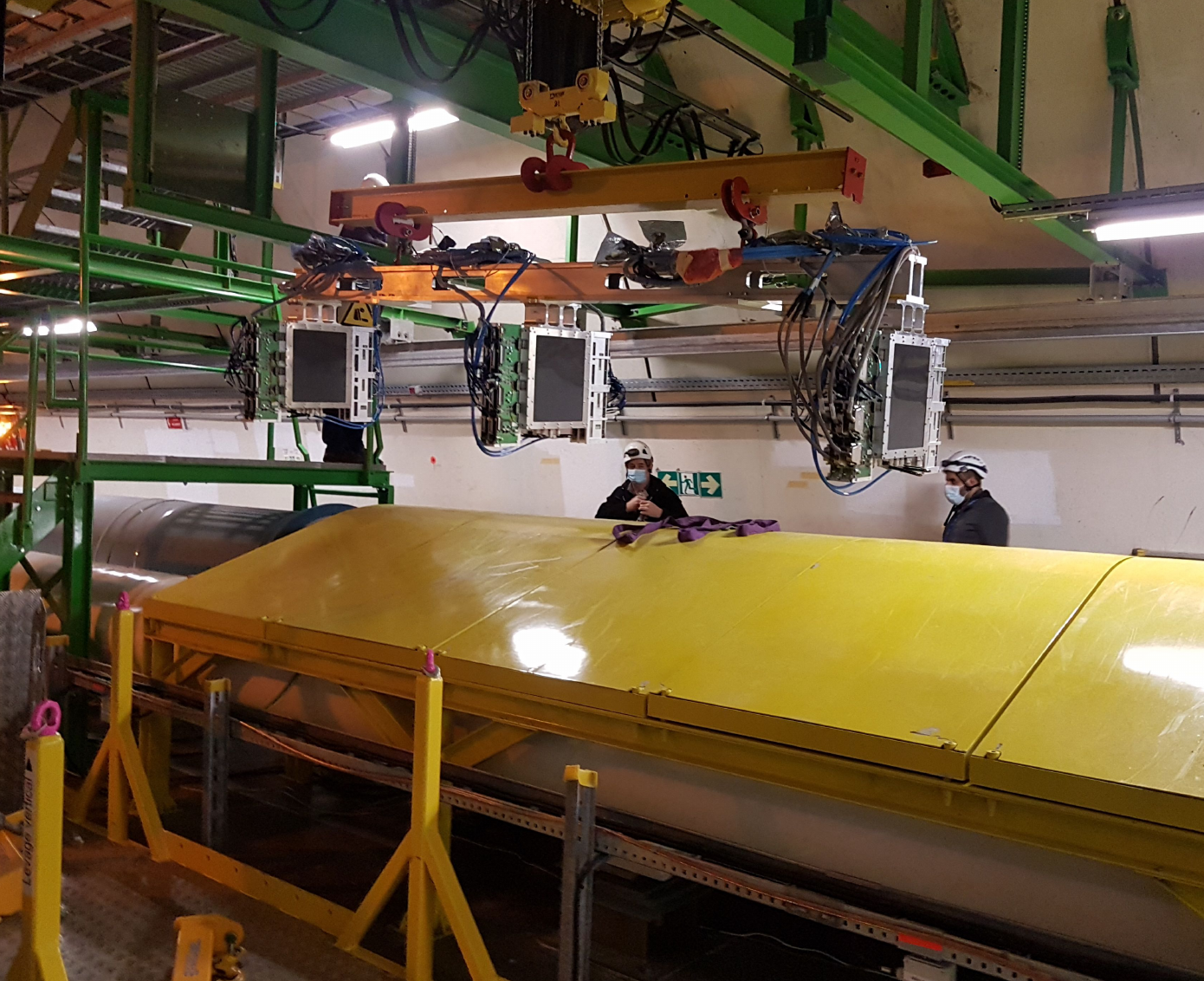}}
\subfloat[\label{f:tracker-CAD}]{\includegraphics[width=0.75\textwidth]{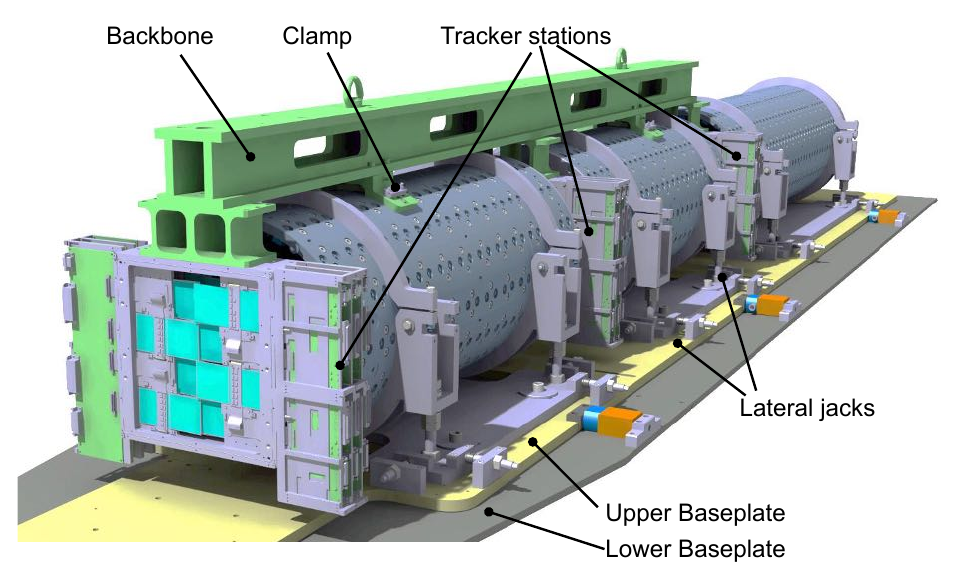}} 
\hfil \\
\subfloat[\label{f:tracker-photo}]{\includegraphics[width=0.64\textwidth]{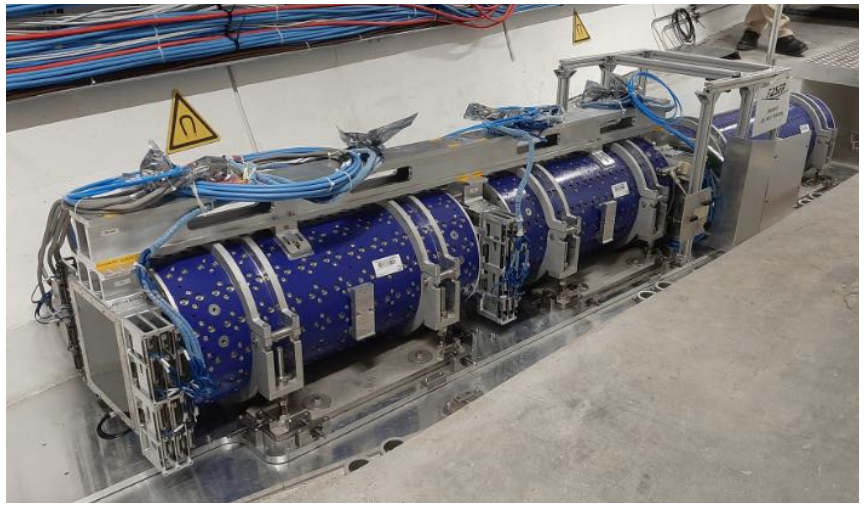}}% 
\caption{(a) Transport of the backbone above the LHC, (b) CAD view of the FASER tracker and (c) photograph of the tracker after installation onto the detector in March 2021.}
\label{f:tracker}   
\end{figure}

The backbone is the primary global mechanical structure of the tracker as it links together the three stations. It allows for an easy handling and transportation (\cref{f:backbone-transport}), and serves as a reference structure for the tracker alignment. The backbone is then supported by the first and second short magnet cylinders via clamps (\cref{f:tracker-CAD,f:tracker-photo}).
The clamp system is a small mechanism that is bolted onto the magnet cylinders and allows some angular movements of the backbone (and thus the tracker stations) with respect to the magnetic field. This angular tuning is the only degree of freedom in view of the tracker station positioning during the survey by the CERN group. One clamp gives precise positioning along the $z$-axis by means of a locating pin, while the second one provides the alignment with respect to the $x$-axis with some dedicated slotted holes.

\section{Powering, cooling and readout system}
\label{sec:equipments}
\begin{figure}[tb]
\begin{center}
\includegraphics[width=9cm]{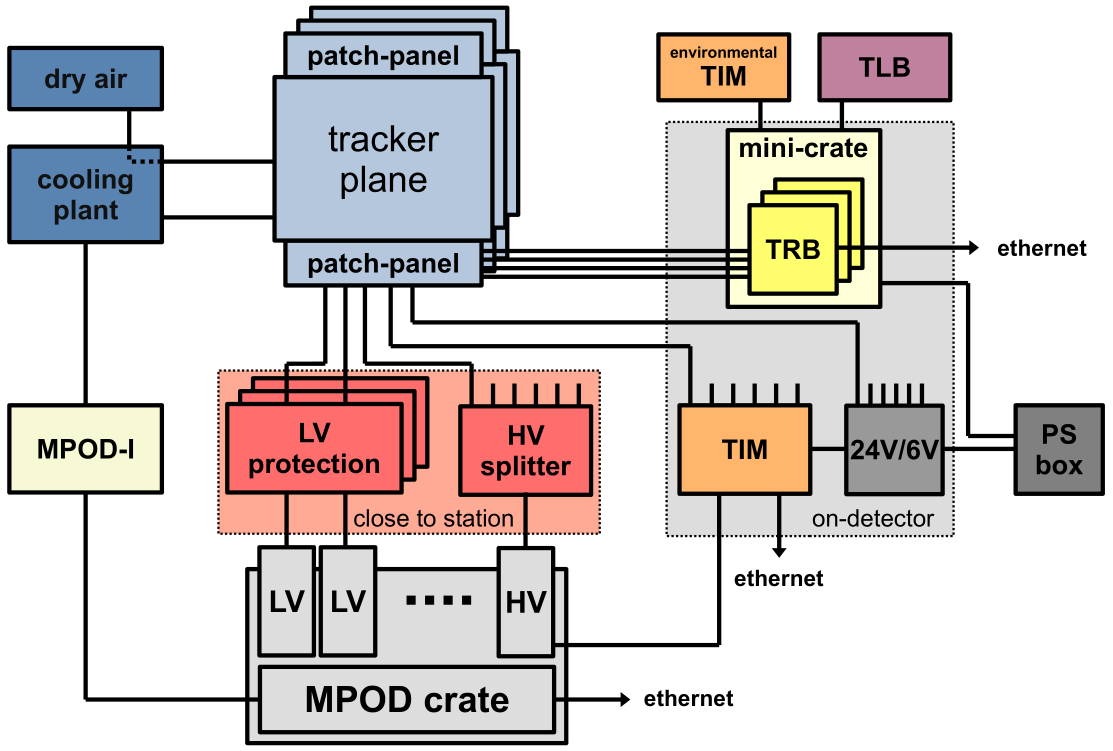}
\caption{Sketch of the FASER detector system. To simplify the sketch, only components connected to one tracker station are illustrated.}
\label{fig:sketch}
\end{center}
\end{figure}

Since access to the FASER location is guaranteed only every two or three months during technical stops of the LHC, it is important that the system is reliable. 
The FASER tracker powering and cooling systems are based on commercial products, and the readout system is based on a custom general purpose I/O (GPIO) board with a dedicated interface card.
Peripheral equipment is designed exclusively for the FASER tracking detector.
A sketch of the detector system is shown in \cref{fig:sketch}, which also includes components for the FASER interlock and detector control system described in \cref{sec:DCS}.

\subsection{Power supply system}
\label{sec:powering}
In order to provide HV and LV to the SCT modules described in \cref{sec:modules}, three 19-inch rack mountable crates called the MPOD LV/HV Power Supply System\footnote{MRAAH2500A2H manufactured by W-IE-NE-R Power Electronics GmbH.} are installed in TI12, which host 3 HV modules\footnote{EHS 84 05p manufactured by Iseg Spezialelektronik GmbH.} and 18 LV modules\footnote{MPV 8008I manufactured by W-IE-NE-R Power Electronics GmbH.} in total. 
For the patch-panels, the readout system (see \cref{{sec:DAQ}}) as well as the interlock and detector control system (see \cref{{sec:DCS}}),
one 19-inch rack mountable box (PSbox) is also installed, which holds fifteen 24V power supplies\footnote{TXL 035-24S manufactured by Traco Power}.

Three types of printed circuit boards, the HV splitter board, the LV protection board and the 24V/6V board, were developed. 
These are mounted on the detector, directly above the tracker stations. 
The HV splitter board divides one HV channel into the four to supply the SCT modules connected to one patch-panel, corresponding to half a tracker plane.
The LV protection board equips an integrated circuit\footnote{LTC4365 manufactured by ANALOGUE DEVICE} which protects against over-voltage possibly induced by radiation in TI12.
The 24V/6V board is powered by the PSbox, and it is used to distribute 24V to the TIM boards, described in \cref{sec:TIM} and 6V to the patch-panels.

\subsection{Cooling system}
\label{sec:cooling-system}

The cooling system was designed and built by the CERN cooling and ventilation group (EN-CV). It consists of two air-cooled water chillers\footnote{HRS030-AF-20MT manufactured by SMC Corporation}: one circulating chilled water to the tracker stations and the second acting as a hot spare. A manifold distributes the chilled water to each tracker station in parallel.

\begin{figure}[tb]
\begin{center}
\includegraphics[width=6cm]{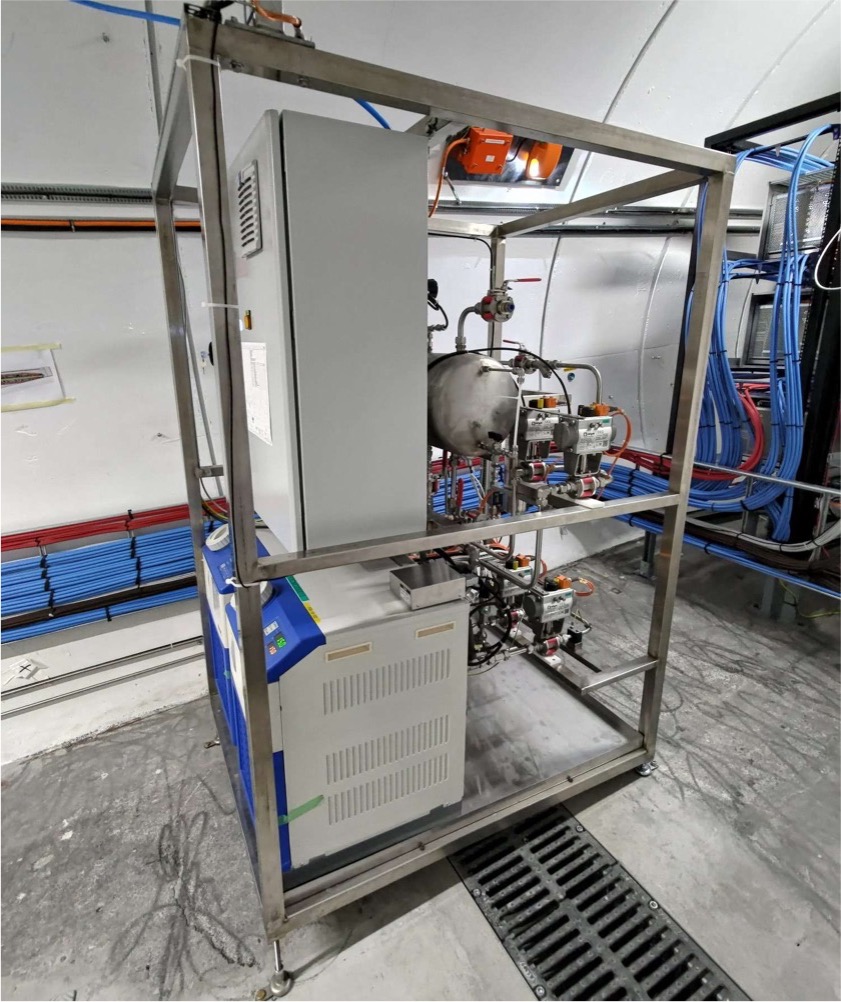}
\caption{Photograph of the cooling system.}
\label{fig:cooling}
\end{center}
\end{figure}

As shown in Fig.~\ref{fig:cooling}, the cooling system is mounted on a single frame together with all instrumentation. An additional water reservoir is also installed, which makes it possible to refill the water tank inside the chiller in an automatic manner.
The cooling capability of each chiller is about 1.8~kW at a \ang{15}C water outlet temperature with $\Delta T=\ang{3}\textrm{C}$ between inlet and outlet temperature. Since one SCT module consumes 6~W, corresponding to an overall power consumption of 450~W for the full tracker, the cooling system is therefore sufficient to regulate the temperature of the tracker stations. 

In case of a failure of the chiller in use for the tracker stations, the cooling circuit is re-routed by controlling valves to the spare chiller to take over the cooling.
Since in normal conditions both chillers are running in the normal state, one connected to the detector and the other in bypass mode, this swap can seamlessly take place without any impact on the operation of the tracker stations. 
If both chillers are not operating correctly, the power supply system is forced to be turned off by the hardware interlock system (\cref{sec:MPODI}).

A dry air system with a dew point of \ang{-40}C coming from compressed air supplied from LHC point 8 is used to flush the tracker stations to avoid condensation on the electronics due to the cooling. In case of a lack of dry air an alarm is triggered and the cooling and detector will be stopped by the hardware interlock system (see \cref{sec:MPODI}).

%===========================================================
\subsection{Readout system} 
\label{sec:DAQ}
\begin{figure}[tb]
\begin{center}
\includegraphics[width=9cm]{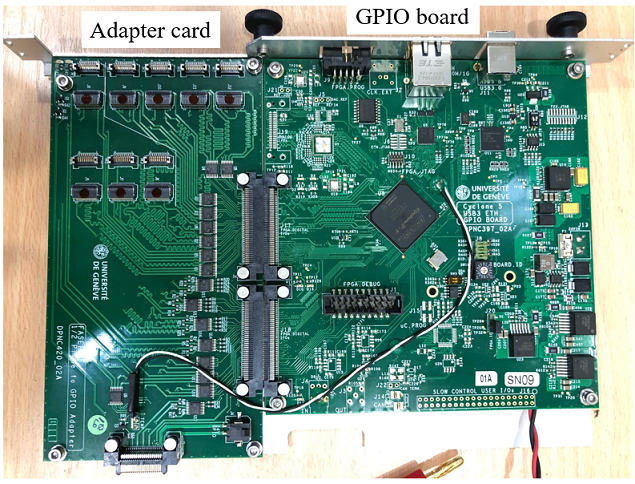}
\caption{TRB consisting of GPIO board and adapter card.}
\label{fig:trb}
\end{center}
\end{figure}

%===========================================================
One Tracker Readout Board (TRB) reads out eight SCT modules corresponding to one tracker plane. Therefore, a total of nine TRBs are used for the three tracker stations in the FASER spectrometer.
The TRB consists of a GPIO board and an adapter card as shown in Fig.~\ref{fig:trb}. The GPIO board was developed as a general readout board centered around a CYCLONE V A7 FPGA (Field Programmable Gate Array). 
The FPGA is driven either by a 40~MHz oscillator on the GPIO board or by an external clock via LVDS input through dedicated two-pins LEMO connectors. The GPIO board is operated with an input voltage of 24~V and provides 5.0, 3.3 and 2.5~V to various active devices either on the GPIO board or to the adapter board. The adapter card is directly attached to the GPIO board and acts as an interface to the patch-panel and the Trigger Logic Board (TLB) \cite{FASER:2021cpr}. The TLB, also a GPIO board, is the central trigger board of the FASER trigger and data acquisition system. There is a single TLB in the system which provides the L1A to all the TRBs. The trigger logic that forms the L1A has as input the signals from the scintillator and calorimeter PMTs, the tracker does not provide an input to the trigger system.

\begin{figure}[tb]
\begin{center}
\includegraphics[width=9cm]{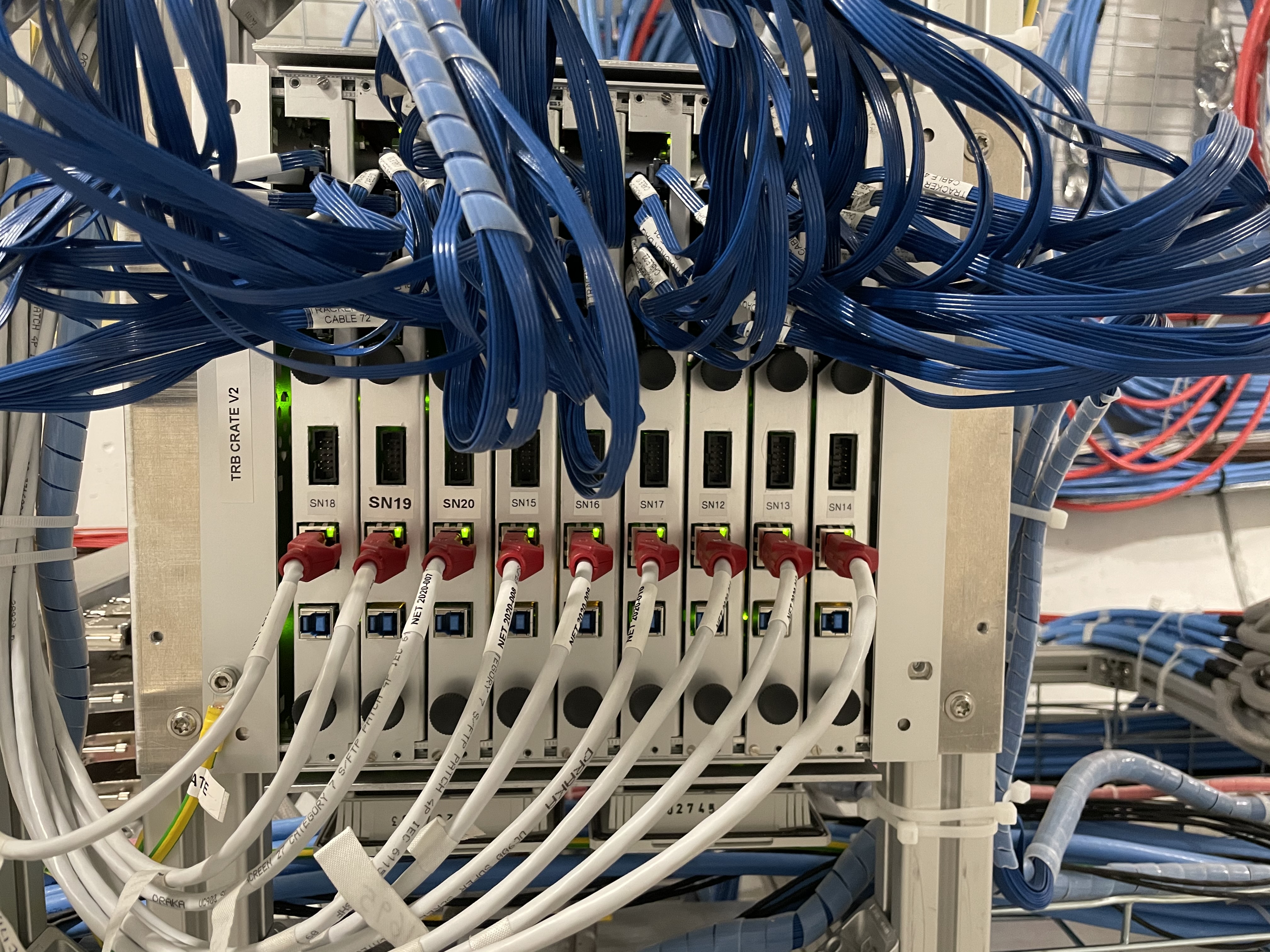}
\caption{TRBs installed in the custom-made minicrate.}
\label{fig:mini-crate}
\end{center}
\end{figure}

For standalone operation during qualification of the tracker planes, the TRB can work independently without any external control and clock signals. On the other hand, in combined operation in the FASER detector, each TRB is controlled by the TLB which provides the 40~MHz clock used by the SCT modules, L1A and bunch counter reset signals. A busy signal is also sent by each TRB to the TLB to hold the trigger during SCT data readout. The nine TRBs are housed on the detector in a custom-made minicrate as \cref{fig:mini-crate} shows, and the minicrate backplane provides 24~V power as well as the TLB signals for the GPIO boards. The TRB is connected to the TLB via RJ45 connector located on the rear side of the minicrate backplane. The GPIO board communicates with a DAQ PC, situated at the surface, via a 1~Gbps Ethernet link.

The software is written in C++ code and manages all operational procedures such as the calibration sequence as described Section~\ref{sec:calibration} \cite{FASER:2021cpr}. The firmware is designed to send low level protocol commands to the ABCD3TA chips on the SCT modules and read out data from them. The high level protocol commands are provided from the PC to the GPIO board, and then the corresponding operation signals are sent to the chips. The data from the chips are stored into First-In First-Out (FIFO) memories in the FPGA and sent to the PC. The software decodes the bit-stream of data from the SCT modules. 
The event and BCID (Beam Crossing Identifier) counters are implemented in the firmware to ensure synchronization between all readout elements (the chips, different TRBs  and TLB). The software in the central DAQ system can send a command to reset the chips as well as the counters and FIFOs in case desynchronization is detected. In addition, if an error is detected in the firmware, it provides error information to the DAQ PC, for example, in the case that the L1 trigger is received while hit data remain in the FIFOs.

The firmware contains four circuit blocks to set the readout timing. Two blocks are used to adjust the clock phase of the 40 MHz input clock with 390~ps steps and  25~ns range, with each four modules on the patch-panel connected to one of these blocks. The other two blocks are used for the circuits in the firmware to latch correctly the output data signals from the modules and compensate for the propagation delay of the cables and the latency in the modules.

The typical data size per plane is expected to be $\sim 212$~Bytes/event. The expected trigger rate at a luminosity of $2 \times 10^{34}$ cm$^{-2}$s$^{-1}$ of 500~Hz corresponds to a required readout bandwidth of at least 106~kBytes/s (848~kbps) with respect to the 1~Gbps Ethernet link. On the other hand, the calibration scans lead to 8.5~kBytes of data per plane per L1A. Since the size of the calibrations data per L1A is much larger than physics data, the bandwidth requirements are dominated by the calibrations.
\section{Interlock and Detector control system}
\label{sec:DCS}
The aim of the safety system of the FASER tracker is to protect the delicate silicon tracker modules from damage under all circumstances. The FASER tracker follows hereby the common approach of a multi-level protection system consisting of a high-level software-based detector monitoring called the Detector Control System (DCS), and a low-level hardware-based interlock system.
The software system is capable of triggering automatic actions that can turn off individual detector components in a controlled way while the hardware interlock system turns off power supplies immediately and acts therefore as the last level of safety.

\subsection{Safety scheme}

\begin{figure}[tbh]
\begin{center}
\includegraphics[width=0.8\textwidth]{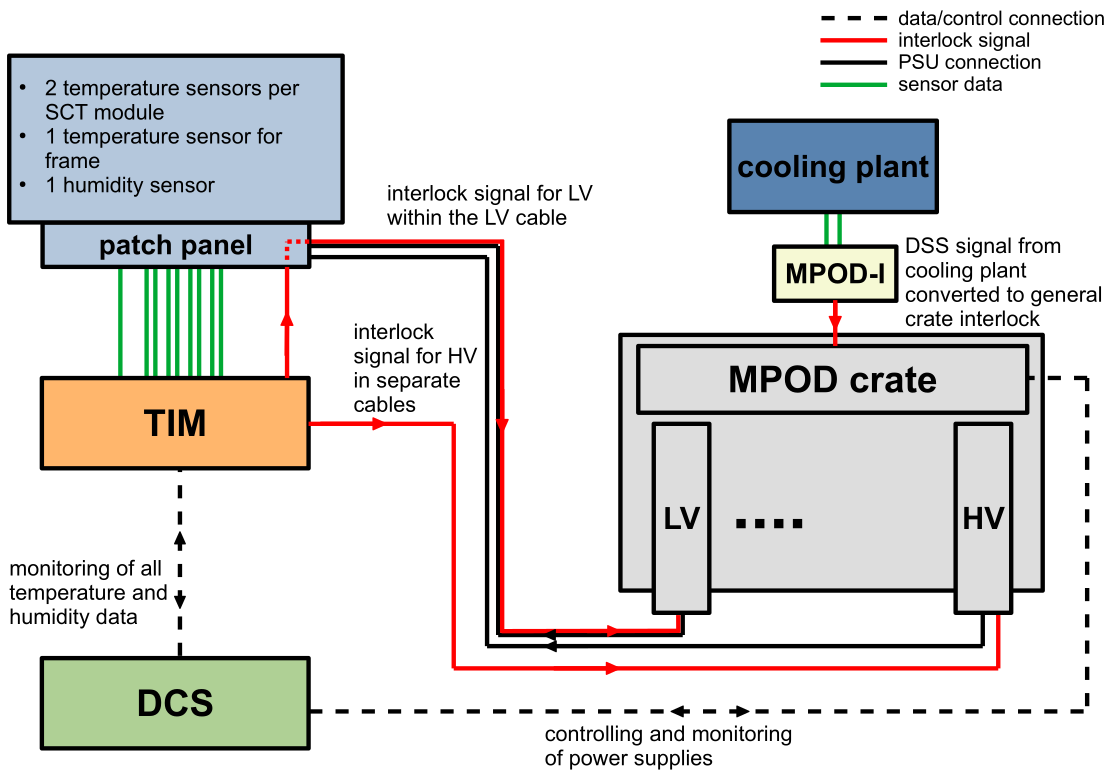}
\caption{Schematic overview of the FASER safety and interlock system. The scheme for one tracker patch panel is depicted, which corresponds to half a tracker plane. A single TIM can serve an entire tracker station (three tracker planes).}
\label{fig:interlock_overview}
\end{center}
\end{figure}

Figure~\ref{fig:interlock_overview} gives an overview of the protection system of the FASER tracker. In the upper left corner, one of the two patch-panels of a tracker plane is sketched. Each patch-panel serves four silicon strip modules and each module is equipped with two temperature sensors (NTC-10k thermistor) that are used for temperature monitoring close to the readout part of the tracker module. In addition, each patch-panel also connects to one temperature sensor (NTC-10k thermistor) that is thermally attached to the mechanical frame inside of the plane as the frame makes the thermal contact to the silicon strip modules themselves. Finally, there is one sensor (HIH-4000) per plane which measures the humidity inside of the tracker station.

All these sensors are read out by the TIM (Tracker Interlock and Monitoring Board). One TIM can receive the signals from a complete tracker station (three planes, six patch-panels). All sensor values are digitized and provided to the software DCS system via an Ethernet connection. Additionally, the frame temperatures of each plane are used as input to an analog comparator circuit which generates an interlock signal for the relevant LV and HV power supplies. The granularity of the temperature interlock can be changed from only a single tracker plane up to the full station by configuring six hardware switches on the TIM.

The interlock signal for HV is provided directly from the TIM to the HV power supplies (one per station) in a separate cable, while the interlock signal for the LV power supplies (one per patch-panel) is propagated to the LV power supply through a dedicated signal line in the LV cable via the patch-panel. The actual interlock logic is the same for HV and LV. The difference in the interlock distribution originates from the differences in the power supplies. While the interlock for the LV power supply is located in the same connector as the output power lines, the HV power supply has a dedicated connector for the channel-wise interlock signal.

In order to prevent the detector modules from electrical damage, each power supply has an independent supervision module that can turn off a channel as soon as it leaves the defined operation range. As an extra level of protection, an additional ASIC for over-voltage protection is placed on the LV protection board in the LV power path.

The final component in the interlock system is the MPOD Interlock (MPOD-I) which receives the Detector Safety Signal (DSS) from the cooling plant and provides a crate-wide interlock to the MPOD crates in case of a cooling plant failure. 

In the following sections the specific hardware components of the tracker safety system are described in more detail.

\subsection{Tracker Interlock and Monitoring} \label{sec:TIM}
\begin{figure}[tbh]
\begin{center}
 \raisebox{-0.5\height}{\includegraphics[width=0.40\linewidth]{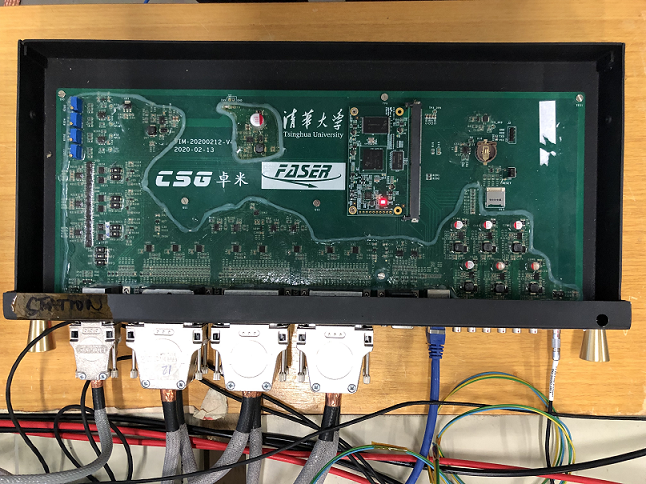}}
 \hfill
 \raisebox{-0.5\height}{\includegraphics[width=0.55\linewidth]{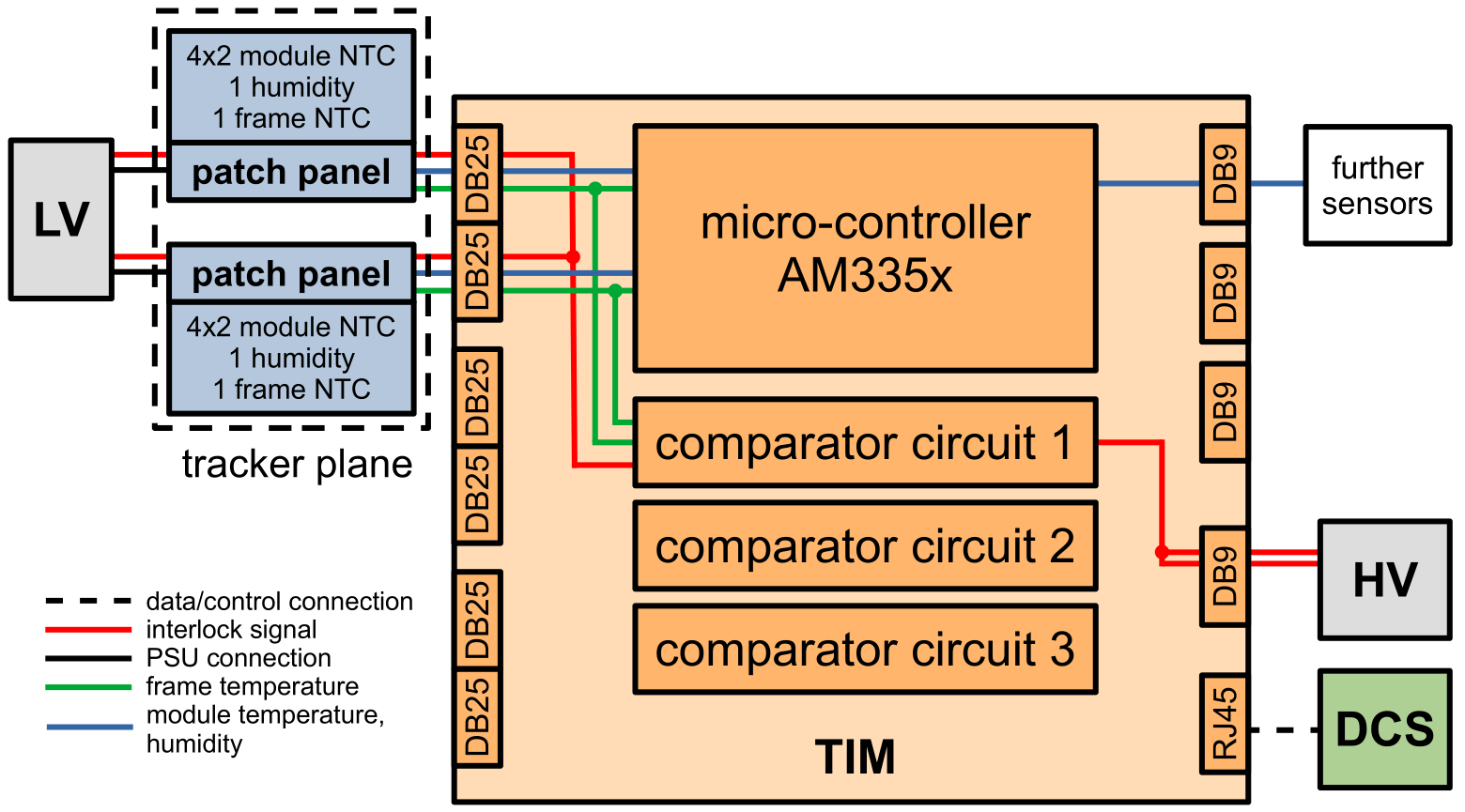}}
\end{center}
\caption{TIM board and corresponding block diagram (signal path shown for only two patch-panels, while six patch-panels can be fed).}
\label{fig:ArchitectureTIM}
\end{figure}

Figure~\ref{fig:ArchitectureTIM} shows a picture of a TIM board as well as its corresponding block diagram.
The core of the TIM is the AM335X micro-controller as well as three independent comparator based interlock circuits.

The TIM can serve up to six tracker patch-panels via DB25 connectors each connector carrying the signal from one frame temperature sensor, $4\times2$ module temperature sensors, and one humidity sensor.
There is one additional DB9 input that allows to connect up to 2 additional environmental temperature sensors.
The latter is currently not used in the FASER trackers setup, but leaves flexibility for further monitoring in the future.

All the mentioned sensors are routed to the microprocessor for digitization and monitoring.
In addition, the frame temperature sensors of patch-panels that belong to the same tracker plane enter into one of the three comparator-based circuits in which a hardware interlock signal is generated if the analog signal of those sensors lies outside of the valid range.
The range itself can be adjusted with potentiometers.
Eventually, two interlock signals are generated for the relevant power supply channels per comparator circuit:
One for the LV power supplies and the other for the HV power supplies.
The interlock is active at the power supply side when no logical high level is applied.

Finally, the TIM features several interfaces for higher level communication: an Ethernet port for the communication with the DCS as well as several serial interfaces for debugging and optional connection to other devices.

\subsection{MPOD Interlock Board}\label{sec:MPODI}
\begin{figure}[tbh]
\begin{center}
 \raisebox{-0.5\height}{\includegraphics[width=0.45\linewidth]{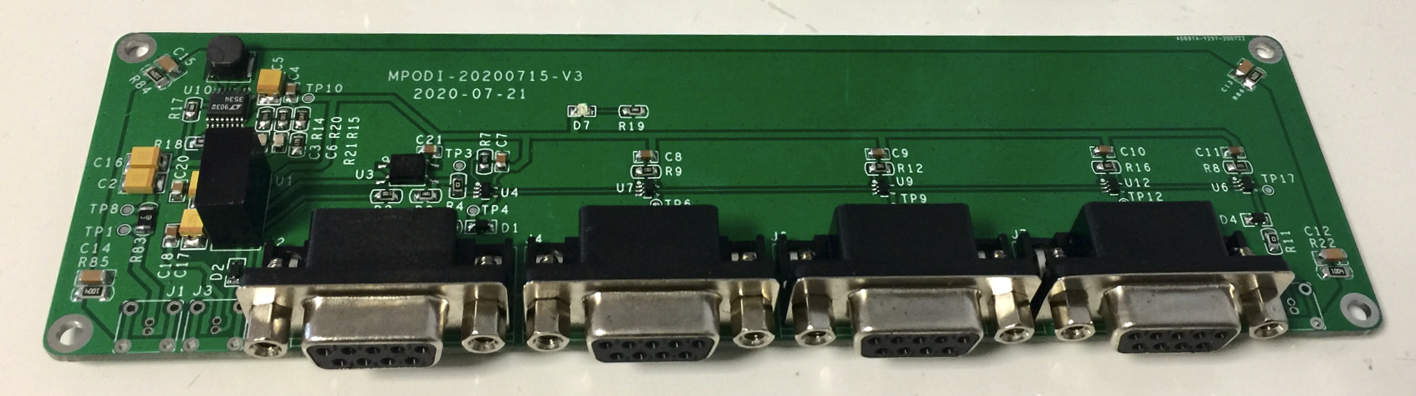}}
 \hfill
 \raisebox{-0.5\height}{\includegraphics[width=0.5\linewidth]{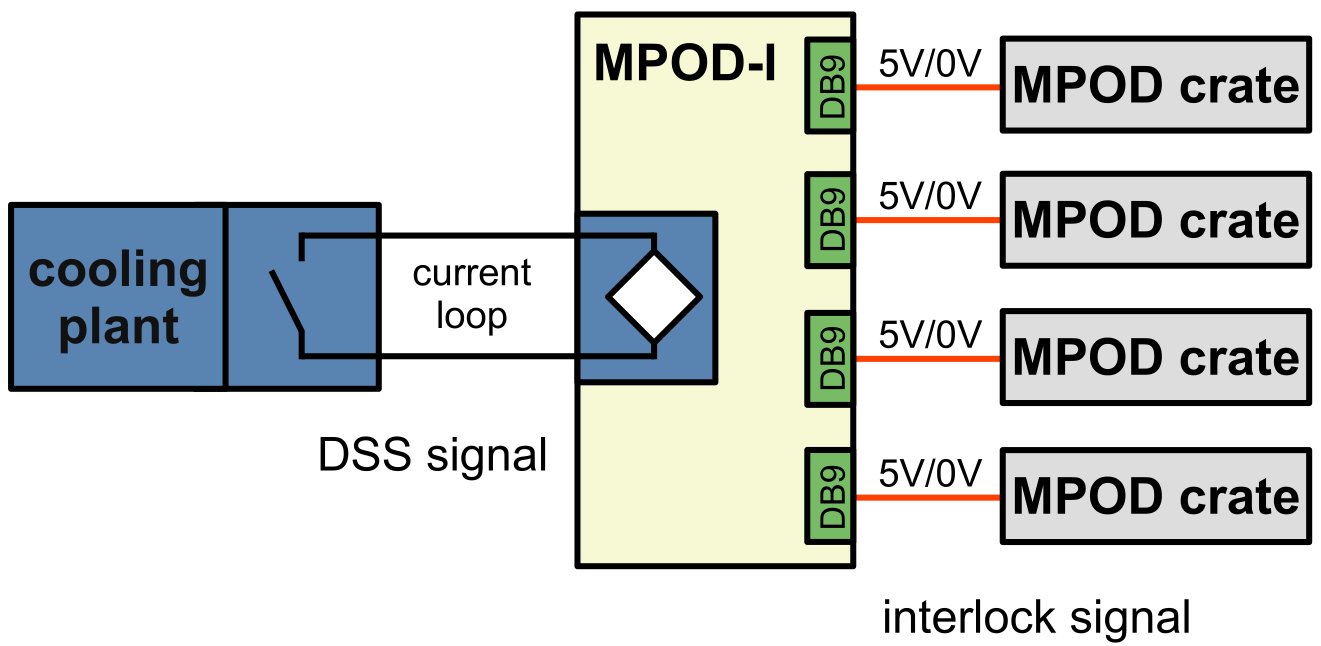}}
 \end{center}
\caption{MPOD-I board and corresponding block diagram.}
\label{fig:MPODI}
\end{figure}

The MPOD Interlock Board (MPOD-I) represents the interface between the DSS signal of the cooling plant and the FASER interlock system.
The board and its corresponding block diagram are depicted in \cref{fig:MPODI}.
The cooling plant PLC controls a loop which is physically interrupted in case of a cooling plant failure (DSS signal).
The MPOD-I checks this loop constantly and translates the DSS signal into a voltage interlock signal which is distributed to up to four MPOD crates housing HV and LV power supply modules, simultaneously.
It is important to note that this interlock signal, in contrast to the interlock signals described earlier, is a crate-wide signal causing the shutdown of all the power supply modules which are part of the crate directly via the crate controller.

\subsection{The DCS system}
The DCS system is the central location where all monitoring data from the TIM boards as well as from the power supplies come together and is stored persistently into a database. Additionally, it provides high level controls of all the power supplies. The software is, therefore, capable of executing automatic actions in case the detector is leaving the operation parameter space indicated by one of the many available sensors. Due to the significantly larger number of sensor readings in the DCS system compared to the interlock system (see \cref{sec:TIM}), a more sophisticated warning and protection scheme can be implemented. It should be noted however that this is not a safety system and does not replace the actual interlock system as it depends on a software process. Details of the DCS system are described in~\cite{FASER:2021cpr}.

\subsection{Limits for DCS automatic actions and hardware interlock}
Table~\ref{tab:interlockDCSlimits} shows the thresholds values of temperature and humidity used for the operation of the tracker safety system. The thresholds for the automatic actions as well as the hardware interlock need to be chosen carefully such that the software-based automatic actions will step in before the actual hardware interlock in case of any abnormal temperature increase. 
\begin{table}[h]
    \centering
    \begin{tabular}{|c|c|c|c|} \hline
        sensor & DCS warning & DCS automatic action & hardware interlock  \\ \hline \hline
        module temperature & $>30.0\,^\circ \mathrm{C}$& $>31.0\,^\circ \mathrm{C}$& -\\ \hline
        plane humidity & $>10\%$ & - & - \\ \hline
        frame temperature & $>23.0\,^\circ \mathrm{C}$ & - & $<5.0\,^\circ \mathrm{C}$ \\ 
                          &                            &   & $>25.0\,^\circ \mathrm{C}$ \\ \hline
    \end{tabular}
    \caption{Thresholds of temperature and humidity to activate the different protection mechanism in the FASER tracker safety system. Note that a lower temperature limit for the hardware interlock is required for NTC temperature sensors in order to interlock on a disconnected NTC ($\mathrm{R}=\infty\,\Omega$).}
    \label{tab:interlockDCSlimits}
\end{table}

\begin{figure}[tbh]
\begin{center}
\includegraphics[width=0.5\textwidth]{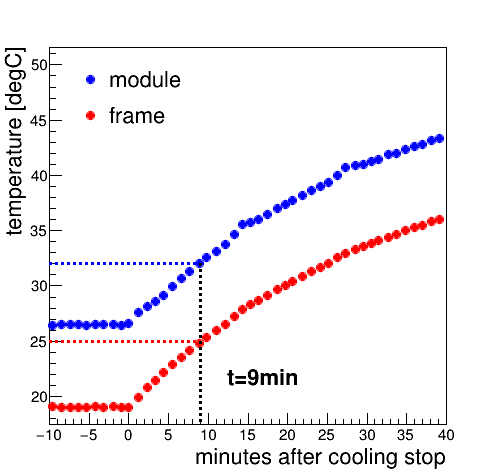}
\caption{Thermal behaviour of frame and module in case of a cooling stop. A special single plane was used for this test and all protection mechanisms were deactivated. The dashed lines indicates the moment when the hardware interlock would become active. The actual module temperature is still well below $35.0\,^\circ \mathrm{C}$.}
\label{fig:thermalbehaviourInterlock}
\end{center}
\end{figure}

In order to verify these limits, a potentially destructive test was performed on a prototype tracker plane.
All protection mechanisms were disabled, the cooling was stopped, and the plane remained fully powered.
Figure~\ref{fig:thermalbehaviourInterlock} shows the thermal evolution measured by a representative frame ($T_{\mathrm{frame}}$) and module temperature sensor ($T_{\mathrm{module}}$).
The condition for the hardware interlock ($T_{\mathrm{frame}}>25.0\,^\circ \mathrm{C}$) was reached after about 9~minutes. The corresponding module temperature at that moment was measured to be $T_{\mathrm{module}}=32.0\,^\circ \mathrm{C}$ and was therefore well below the required maximum module temperature of $35.0\,^\circ \mathrm{C}$. The threshold for the automatic DCS actions are adjusted in such a way that they are sufficiently far away from the normal operation point, but low enough in order still trigger before the hardware interlock under normal circumstances.

%\begin{figure}[ht!]
%\begin{center}
 %\includegraphics[width=0.6\linewidth]{figure/data_trans.png}
 %\end{center}
%\caption{Data Transformation from TIM to DCS.}
%\end{figure}

%\begin{figure}[ht]
%\begin{center}
% \includegraphics[width=0.9\linewidth]{figure/MPODI_logic.png}
% \end{center}
%\caption{MPODI board and schematic of main functions.}
%\end{figure}

% [TBD, Zhen \& Candan]

\section{Tests during construction and commissioning of the FASER tracker}
\label{sec:commissioning}
The FASER tracker elements were tested at each stage of the construction. 
The electrical performance and behavior of the silicon sensors as a function of the applied bias voltage were investigated for the single SCT modules, individual tracker planes and full tracker stations on the surface before installation of the tracker stations into the FASER experimental site. 
In addition, metrology was performed for the layers and stations. After the installation, the performance was also tested using cosmic rays and random triggers deploying the central DAQ system of the FASER experiment. In this section, the test setup, procedure and results of the thermal and electrical tests are described. 

\subsection{Test setups during testing on surface}
\label{sec:comm-setup}
Three test setups were installed to perform the required measurements at the different stages, namely tests of a single module, plane and station. 

The single module test was done with the readout system developed at Cambridge University \cite{Keizer:2018nju}. It was used to evaluate the electrical performance of a single module. A chiller was used to cool down an environmental enclosed box to keep the module below 30$^{\circ}$C. The number of functional strips, the noise value and dependency on applied bias voltage (high-voltage behaviour) of the silicon sensors were verified and compared to the results in the module production of the ATLAS SCT detector~\cite{Abdesselam:2006wt}.

\begin{figure}[th]
\centering
\includegraphics[width=0.65\textwidth]{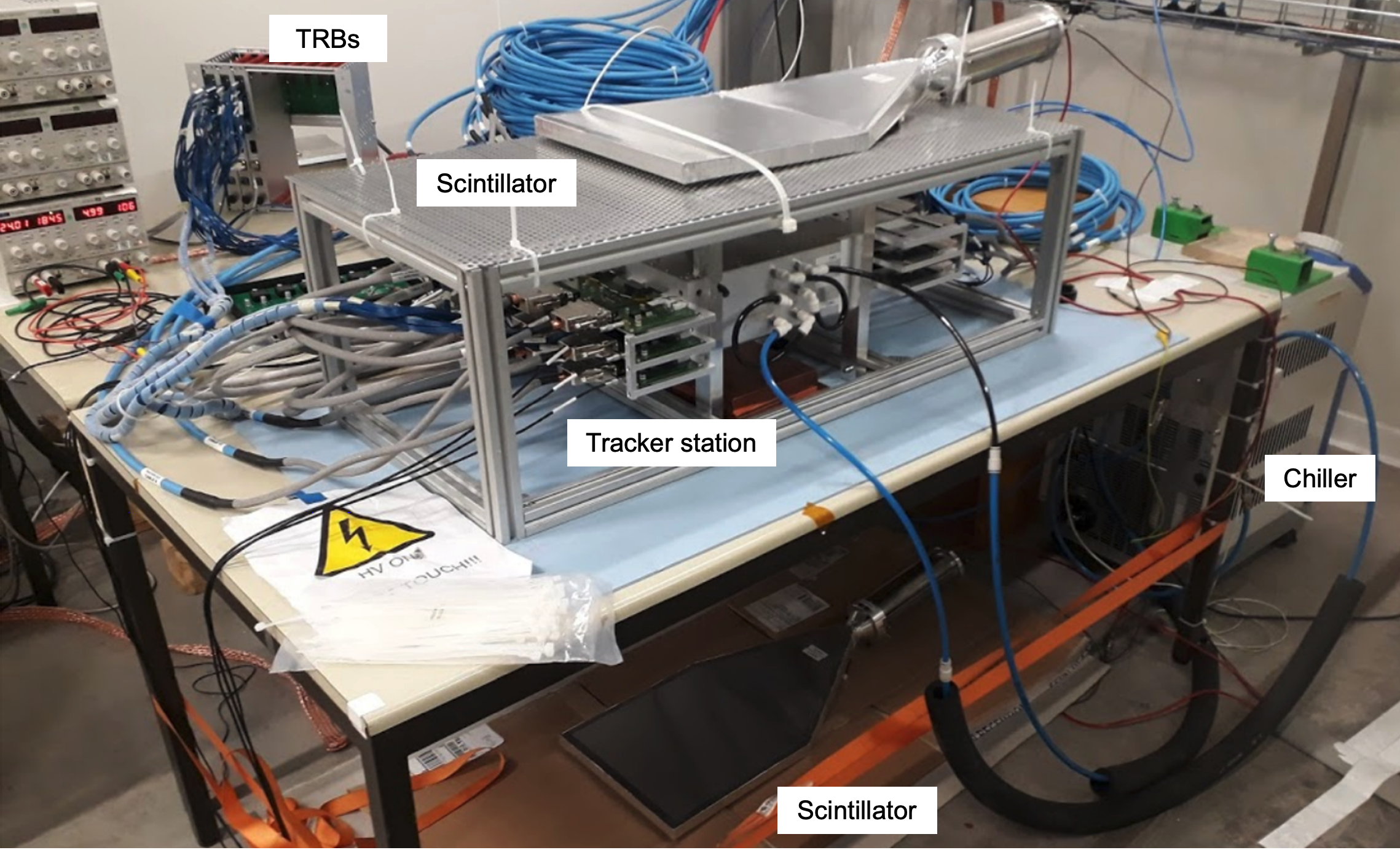}
\caption{The test setup for the station commissioning on surface.}
\label{f:setup}
\end{figure}

The individual planes and stations were qualified on surface in test-stands which used equipment later used in the FASER experiment. Both the interlock and monitoring of the temperature, voltage and current were handled by the TIM unit (see Section~\ref{sec:TIM}). This allowed the DCS information to be monitored live and to be archived into a database for evaluating the detector performance. The same powering and DAQ systems, cables and calibration software as used for the experiment were used during the testing on surface.
This allows to compare the electrical performance at different stages of the assembly and installation as well as from different data taking periods in the tunnel. The surface commissioning with the components used in the real experiment made it possible to test their operation and investigate the long term behaviour early on. 
The operation of the different tracker planes and stations was conducted with specific finite state machines which took the modified detector mapping on the surface compared to the one in the tunnel into account.
Figure~\ref{f:setup} shows a picture of the test setup with its main components for station commissioning.

\subsection{Test procedure}
\label{sec:thermalproc}
After installation of the plane or station under test into the test setup, thermal and electrical measurements were conducted. The thermal measurements with NTCs on the frame of the planes and SCT modules are performed without powering the SCT modules and with powering the modules before and after configuring the ABCD3TA chips.
During the electrical characterization, the leakage current of the sensors was measured. The LV and its current consumed were also investigated after powering up and configuring the chips. 
Further important parameters measured in the electrical calibration sequence are the number of non-functional channels, so-called masked channels, the noise, the gain and the noise occupancy of each module after trimming the threshold offset of the ABCD3TA chips as explained in Section~\ref{sec:calibration}. The same test procedure was used when commissioning the tracker planes and the stations to allow for comparisons between the collected data.
The planes and stations were powered for at least 24 hours to investigate their long-term stability and behaviour of the leakage current in the silicon sensors.

The performance of each module was assessed during each step of the station assembly procedure from individual module testing, to mounting in a plane, and finally in a station.
The plane and station were signed-off, considering a list of the required parameters such as thermal stability, less than 0.5\% masked strips that are identified as either very noisy or dead, noise and gain values comparable to the single module testing.
After installing all three stations into the FASER experiment in TI12, the same thermal and electrical tests were conducted and the test results were compared to the surface commissioning data. A general agreement was found as shown in more detail below.

\subsection{Results of thermal tests}
\label{sec:thermaltests}
\begin{figure}[th]
\centering
\includegraphics[width=0.95\textwidth]{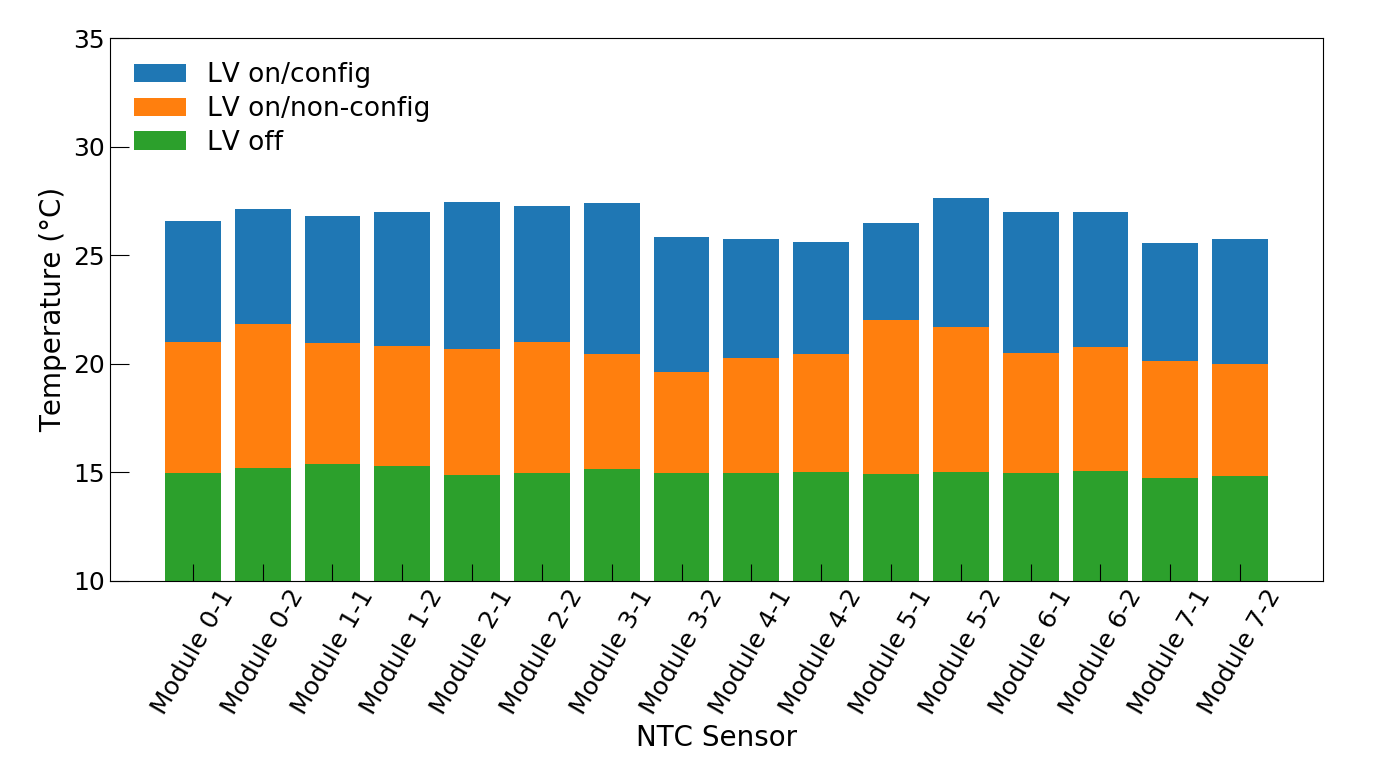}
\caption{Thermal test results of one plane in the first station during surface commissioning. Temperature values of the SCT module NTCs for power-off (green), after powering the modules (orange) and after powering and configuring (blue).}
\label{f:thermST1}
\end{figure}

Figure~\ref{f:thermST1} shows the temperature of the two NTC sensors on each module in one plane of one station, which was measured during surface commissioning of the first station. The temperature values are listed for power-off (green), after powering the modules (orange) and after powering and configuring the chips (blue). The values without powering correspond to the coolant temperature of 15$^{\circ}$C. The maximum values of around 27$^{\circ}$C are reached after powering and configuring the modules. The module temperatures vary slightly in a pattern along the flow of the cooling loop. It leads to the behaviour that typically the modules in the corners of the plane (modules 0, 3, 4, and 7) are the coldest ones. 

All the temperature values are well below 35$^{\circ}$C which is the maximum temperature of the module. The results are consistent within 1$^{\circ}$C with those in the plane commissioning.
A similar thermal performance was obtained for the other two stations and after installation into TI12. 

\begin{figure}[th]
\centering
\includegraphics[width=0.65\textwidth]{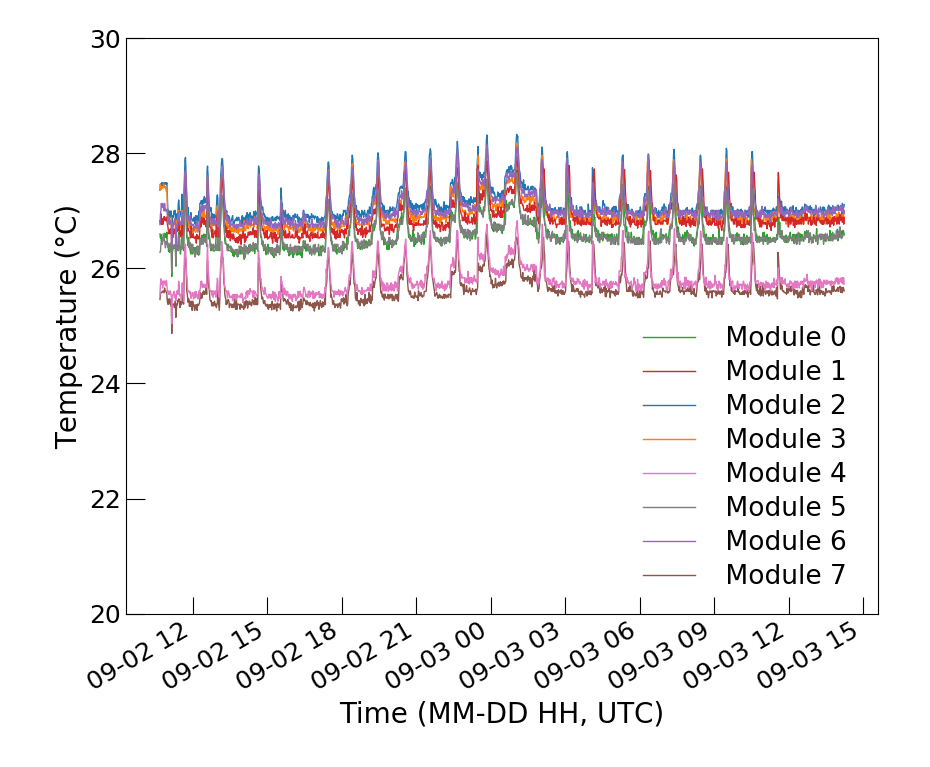}
\caption{Temperature values of one of two NTCs on the SCT modules of plane 5 during a scan sequence for 24 hours in plane commissioning.}
\label{f:SCTNTCs}
\end{figure}

As an example of the temperature stability of the tracker planes, Figure~\ref{f:SCTNTCs} shows the temperature measured with one of two NTC sensors
on each module in plane 5 during a period of 24 hours covering one calibration sequence.
The analogue and digital currents consumed by the ABCD3TA chips fluctuate during the calibration, causing a slight variation of the module temperature, however this never exceeds 29$^{\circ}$C. When the module is powered off, the temperature decreases to the coolant temperature of 15$^{\circ}$C. 
It is confirmed that the temperature was stable during nominal operation. In addition, the dew point inside the planes is kept around \ang{-40}C, which is well below the coolant temperature during the commissioning.

\subsection{Calibration procedure} 
\label{sec:calibration}

The readout of the SCT barrel modules is binary, meaning there is only a hit or no-hit information depending on whether the current pulse generated by a charged particle passing through the silicon sensor is above or below the threshold set. Since the analog information of the signal pulse is not recorded, a good calibration of the ABCD3TA chips on the SCT module is essential. For this purpose, the chip contains an internal calibration circuit that enables to simulate a hit pulse in a strip. Each channel of the readout ASIC has a \SI{100}{\femto F} calibration capacitor connected to its input. An internal 8-bit DAC and chopper circuit are used to generate a voltage pulse that is sent to the calibration capacitor, thus by-passing the sensor strips. The amplitude of the calibration pulse can be set in the range from 0 to \SI{160}{\milli\volt} (corresponding to input charges up to \SI{16}{\femto\coulomb}), and its delay with respect to the clock phase can be adjusted in 64 steps within 50\,ns. 

\begin{figure}[th]
\centering
\includegraphics[width=0.65\textwidth]{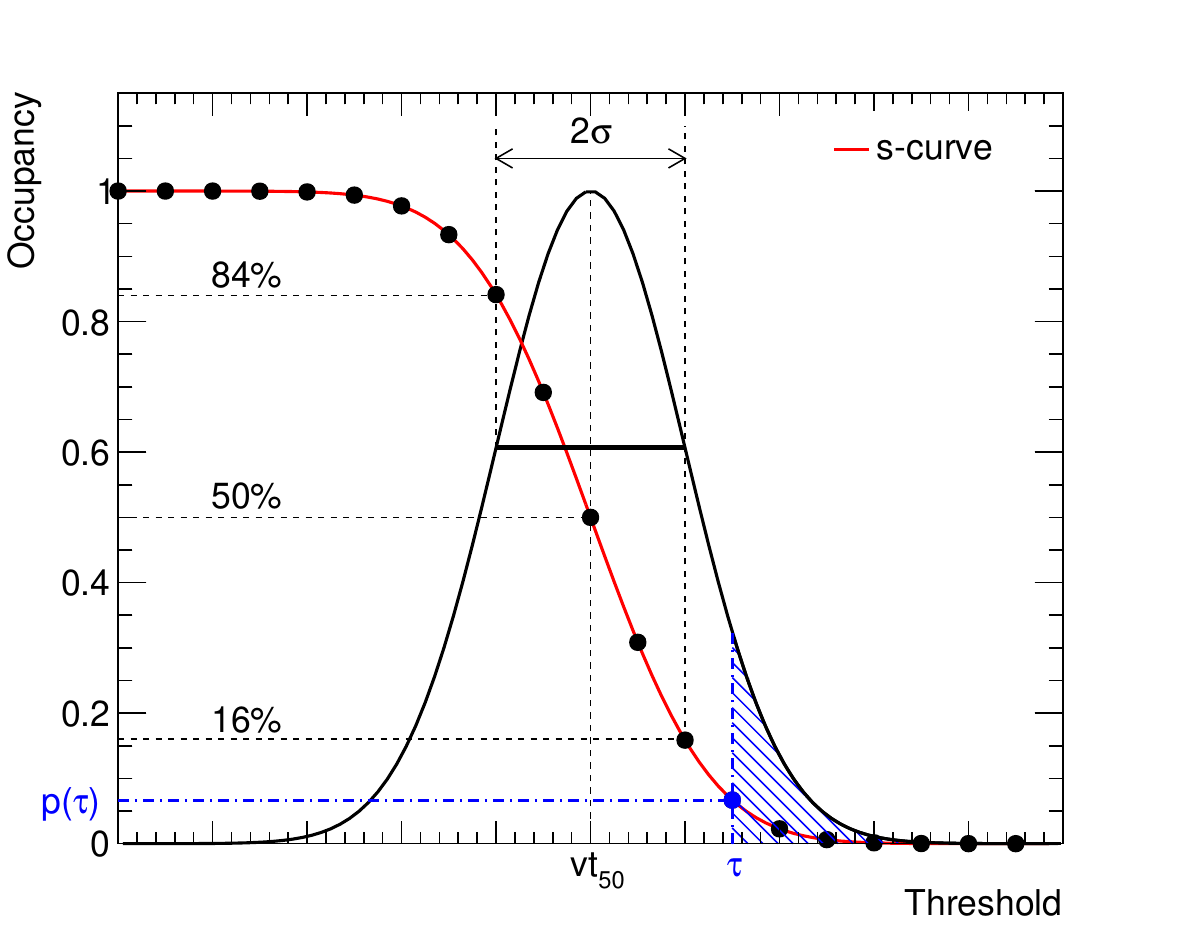}
\caption{Illustration of a threshold scan, in which a certain number of calibration charges are injected at each threshold point. The input signal is convoluted with electronics noise, described by a Gaussian distribution function. For a given threshold $\tau$, the hit occupancy is the fraction of signals above threshold (dashed area). The resulting s-curve, essentially a complementary error function, corresponds to the probability $p(\tau)$ that the signal is above the threshold $\tau$.}
\label{f:scurve}
\end{figure}

The calibration of the chips mostly relies on so-called {\em threshold scans} (see~\cref{f:scurve}), in which the discriminator threshold (differential voltage generated from an internal 8-bit DAC in the range from 0 to 640 mV) is varied in discrete steps and a set of well defined fixed-amplitude calibration charges is sent at each step. 
The hit occupancy, defined as the fraction of injected signals above threshold, is computed for each readout channel at each threshold point. 
Since the signal amplitude is convoluted with Gaussian electronics noise, the hit occupancy does not follow an ideal step function along the threshold setting. 
The distribution behaves as the so-called {\em s-curve}. 
The threshold at which the occupancy is 50\% is called the $vt_\text{50}$ point and corresponds to the input charge amplitude. 
The width of the Gaussian distribution (determined from a fit to the s-curve function) is the noise at the discriminator output.

The calibration procedure implemented for the FASER tracker largely follows that established by the ATLAS SCT collaboration~\cite{SCT-calib}, which has been extensively used during past years to perform the electrical characterization of the SCT modules. Typically, several tests / scans are run in sequence, and the chip parameters are updated along the sequence. A typical calibration sequence is listed below.

% charge injection + L1A
% standlonw C++ code, work in progress for integration into the main run control

\begin{enumerate}

\item {\em Mask-scan}. This test aims at determining two sorts of main defects, {\em i)} dead / non-responsive channels, and  {\em ii)} very noisy channels. This is achieved by setting a very low (high) threshold, then  sending a number of trigger signals (without charge injection) to read the data stored in the digital pipelines, and finally checking how many channels are below (above) threshold. The strips with very low (high) occupancy at the low (high) threshold setting are identified as dead / non-responsive (noisy channels). The identified defective channels are disabled by tagging them in a  dedicated in-chip mask register. 

\item {\em StrobeDelay scan}. This scan is used to determine for each chip the optimum delay between the calibration charge and the clock. The calibration delay is varied in its full range and at each step the occupancy per channel is computed after charge injection. Then, the occupancies are projected to the calibration delay axis and a fit is carried out. An optimum delay per  is calculated to be in the plateau of maximum occupancy.  

\item {\em Three-Point-Gain}. This test allows to estimate for each channel the gain and noise of the preamplification stage. Threshold scans are performed for three different input charges. For each strip a linear fit of the $vt_\text{50}$ thresholds versus injected charge is carried out to obtain the gain (slope) in mV/fC and threshold offset (intercept) in mV. The Equivalent Noise Charge (ENC) at the discriminator input is calculated as the standard deviation of the s-curve fits in mV ($\sigma_{vt_\text{50}}$) divided by the gain. The ENC is evaluated with 2~fC of input charge, which is the value used to study the performance of the ATLAS SCT~\cite{ATLAS:2014bdf, ATLAS:2021zxb}.

\item {\em Trimming}. This scan aims at correcting for the threshold dispersion inside the chip. A 4-bit resolution DAC (TrimDac) allows to set individual channel threshold corrections. The TrimDac has four possible ranges, as the ABCD3TA chip was conceived to operate in the LHC environment in which the threshold spread increases with time due to accumulated radiation damage to the sensors and electronics. In the trimming procedure threshold scans are performed for various settings of the TrimDac (ranges and step values). In our calibration routine, four DAC settings are used. %Given the excellent linearity of the DAC, four out of the sixteen possible steps per range are used. 
Using a linear fit to the $vt_\text{50}$ as a function of trim value, a channel is flagged as ``trimmable'' if its offset can be corrected with respect to a given threshold target. The trimming settings are those that correspond to the minimum TrimDac range and the minimum threshold target (within that range) for which a maximum number of channels are trimmable. This is done on a module-by-module basis to achieve a good threshold uniformity across all channels of a given module.

\item {\em ResponseCurve}. With this test an accurate threshold-to-charge relation is obtained. Threshold scans are performed for ten input charges, from \SI{0.5}{\femto\coulomb} to \SI{8}{\femto\coulomb}. After the corresponding s-curve fits, for every channel the $vt_\text{50}$ as a function of the input charge is fit to a first-degree polynomial with an exponential correction term to account for small non-linearities at very low and high injected charges, $\tau = p_2 + p_0/(1+e^{-q/p_1})$, where $p_0$, $p_1$ and $p_2$ are the fit parameters, and $\tau$ and $q$ are the threshold in mV and fC, respectively. The average parameters for each individual ASIC are obtained to determine the mV to fC conversion.

\item {\em NoiseOccupancy}. Although this test is formally not part of the chip calibration (as no calibration parameters are derived from it and it is not used to mask any additional strips), it is typically performed to assess the goodness of the above-mentioned procedure. The noise occupancy (NO) is defined as the probability for a strip to give rise to a hit only due to noise. This typically occurs when fluctuations at the discriminator input exceed its threshold. For the ATLAS SCT modules the noise occupancy per strip is specified to be less than $5\times 10^{-4}$ at \SI{1}{\femto\coulomb} threshold and the nominal operating temperature. The NO is determined by performing a threshold scan without any input charge. The number of triggers sent is increased progressively as the threshold is raised.

\end{enumerate}

\subsection{Results of electrical and calibration tests}

The calibration scans described in Section~\ref{sec:calibration} were repeated during individual plane commissioning, station commissioning, and commissioning \emph{in situ} after installation in TI12 as described in Section~\ref{sec:comm-setup}. The aim of the calibration is to achieve uniform threshold distribution, high hit efficiency ($> 99\%$) and low noise occupancy ($< 5\times10^{-4}$) at the nominal operating threshold of 1\,fC. 
In this sub-section, electrical calibration results are shown and compared between single module testing, plane testing and station testing. Unless otherwise stated specially, the results of the station test are obtained after the installation in TI12 during March/April 2021, however they are consistent with the station results obtained on the surface.

\begin{figure}[th]
\centering
\includegraphics[width=0.65\textwidth]{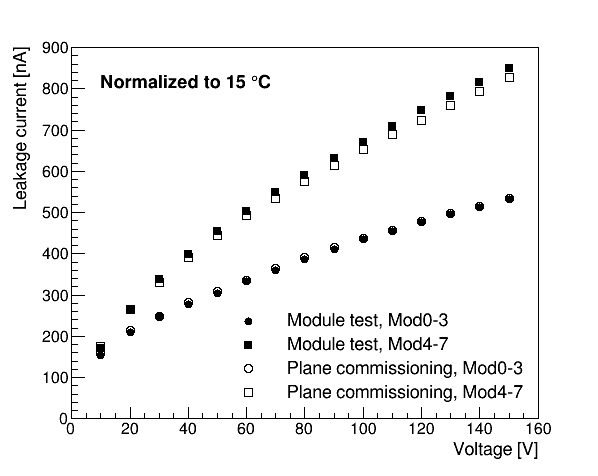}
\caption{Comparison of leakage current measurements during single module testing and plane commissioning.}
\label{f:IV}
\end{figure}

The leakage current of the silicon sensors was measured as a function of the applied bias voltage in the tests of a single module, plane and station. The curves are generally in fair agreement between the different stages. 
Figure~\ref{f:IV} shows a comparison of the leakage current measured during single module commissioning and the plane commissioning. The individual module leakage currents are typically in the range $100$--$800$~nA at~$15 ^{\circ}$C. Since in the plane four modules are connected to one HV channel, the comparison is made by summing the individual module leakage current measurements for the four modules in a half a plane (either Module 0-3 or 4-7) (scaled to 15~$^{\circ}$C) and comparing to the measurement in the plane itself (channel 0 or channel 1). Similar agreement was also found when comparing results of plane and station commissioning.

In order to achieve a high hit efficiency, the number of masked strips -- those identified as either dead or very noisy during the mask scan -- in each module was carefully monitored during the tests for a single module and the plane and station commissioning. The modules with lower numbers of masked strips were selected to be mounted at one of the four inner module positions which are inside the central region of the magnet acceptance, as demonstrated in Figure~\ref{f:station-material}. In addition, the station with the highest quality -- that with lowest number of the masked strips in this region -- was chosen to be placed at the front (upstream) of the spectrometer while the station with the lowest quality is located at the back (downstream).
An electron and positron pair from the decay of a dark photon is most collimated at the upstream station, therefore, the station with the best performance was selected for that position.
Table~\ref{tab:masked-strips} shows the number of masked channels observed during the \emph{in situ} commissioning after the installation. The four modules in the inner region have a maximum of 0.08\% of masked channels. Even in the outer region, this is less than $\sim0.3$\%.

\begin{table}[h!]
    \centering
    \begin{tabular}{|c|c|c|c|} \hline
        Station & Inner region & Outer region \\ \hline
        Station 1 (upstream) & 0.02\% & 0.06\% \\
        Station 2 (middle) & 0.03\% & 0.04\% \\
        Station 3 (downstream) & 0.08\% & 0.31\% \\
        \hline
    \end{tabular}
    \caption{Fraction of masked channels in each tracker layer for four modules in the inner (outer) region.}
    \label{tab:masked-strips}
\end{table}

\begin{figure}[th]
\centering
\subfloat[\label{f:gain}]{\includegraphics[width=0.49\textwidth]{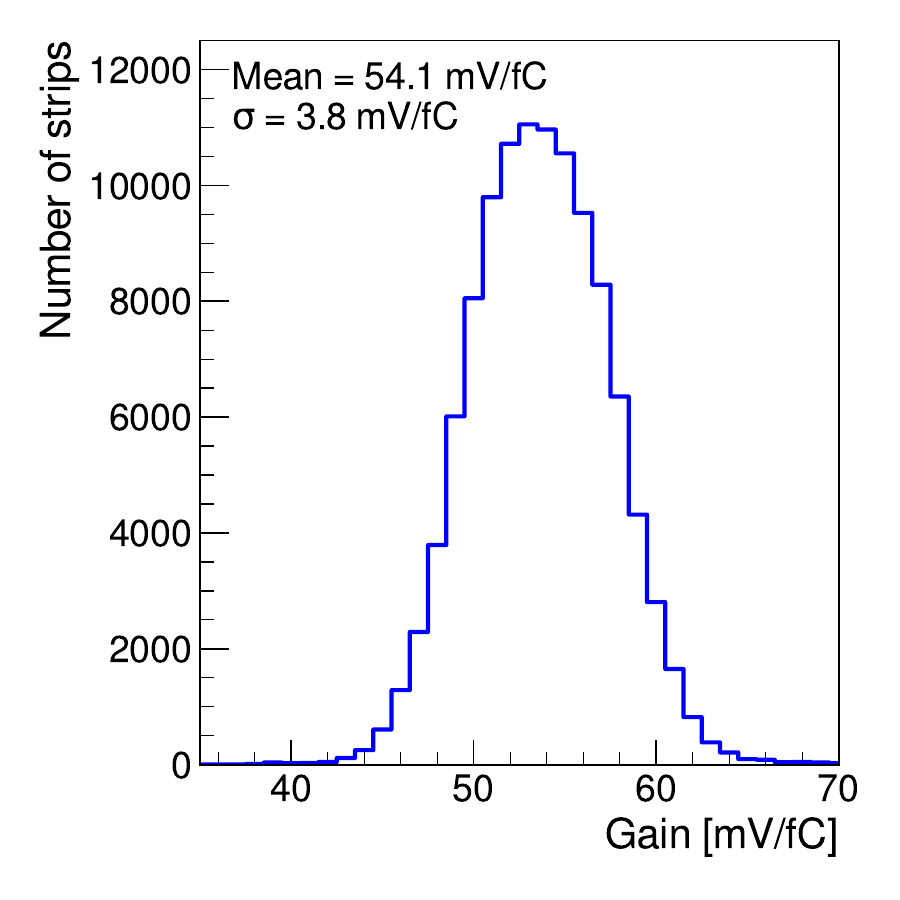}}
\hfil
\subfloat[\label{f:response-curve}]{\includegraphics[width=0.49\textwidth]{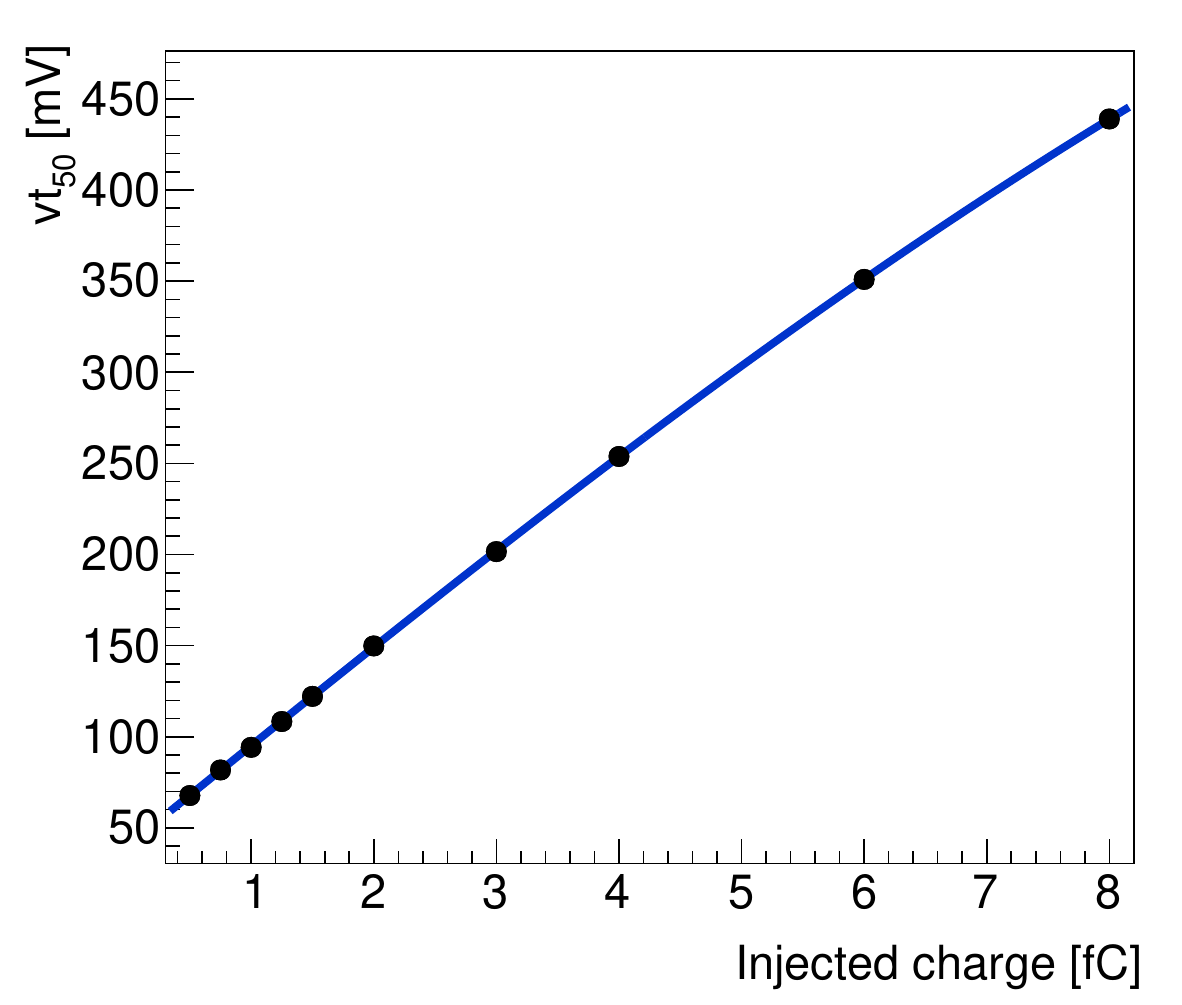}}
\caption{(a) Distributions of the strip gain in each station. (b) Example response curve fit for a single strip. }
\label{f:gain-rc}
\end{figure}

The gain measured in the three-point-gain measurements are shown in Fig.~\ref{f:gain}. 
The average over all the strips in all stations is 54~mV/fC, which is in good agreement with the $\sim 55$ mV/fC expected from the module specification~\cite{Abdesselam:2006wt}. 
The dependence of the gain over a larger range of injected charges is tested in the response curve scan. 
Figure~\ref{f:response-curve} shows the typical result of the response curve scan for a single strip.

\begin{figure}[th]
\centering
\subfloat[\label{f:trim_before}]{\includegraphics[width=0.49\textwidth]{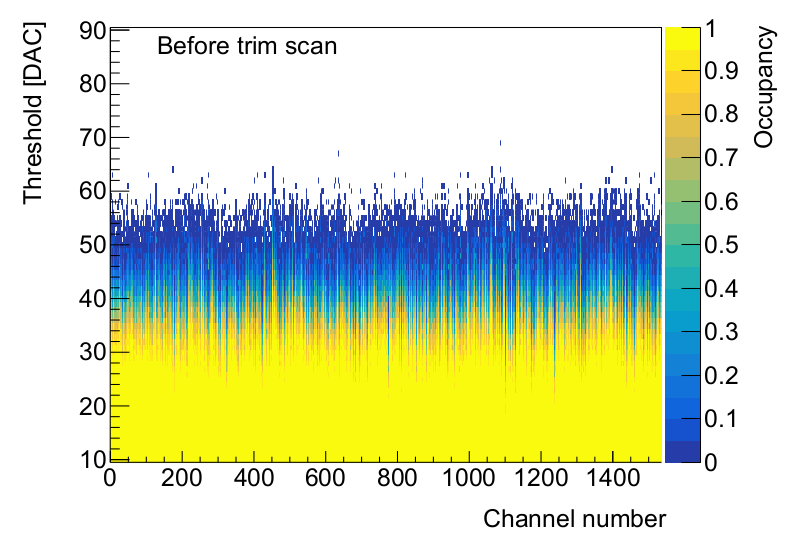}}
\subfloat[\label{f:trim_after}]{\includegraphics[width=0.49\textwidth]{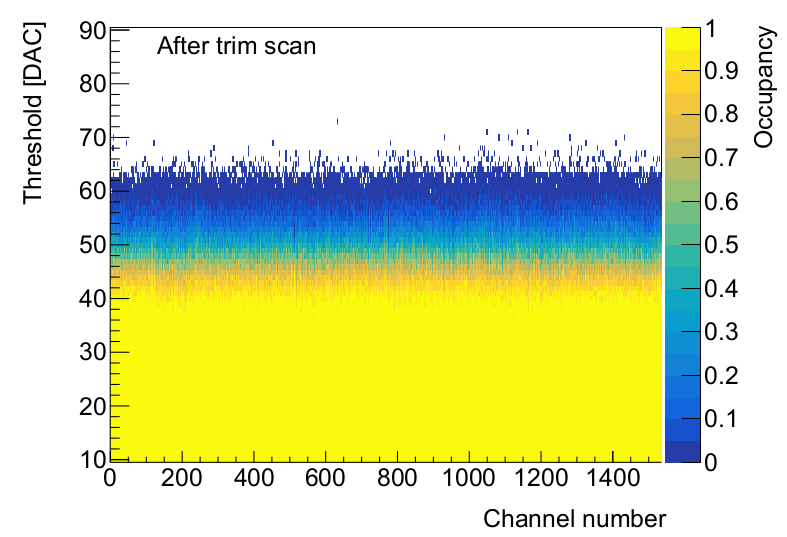}}\\
\subfloat[\label{f:trim_before_vt50}]{\includegraphics[width=0.49\textwidth]{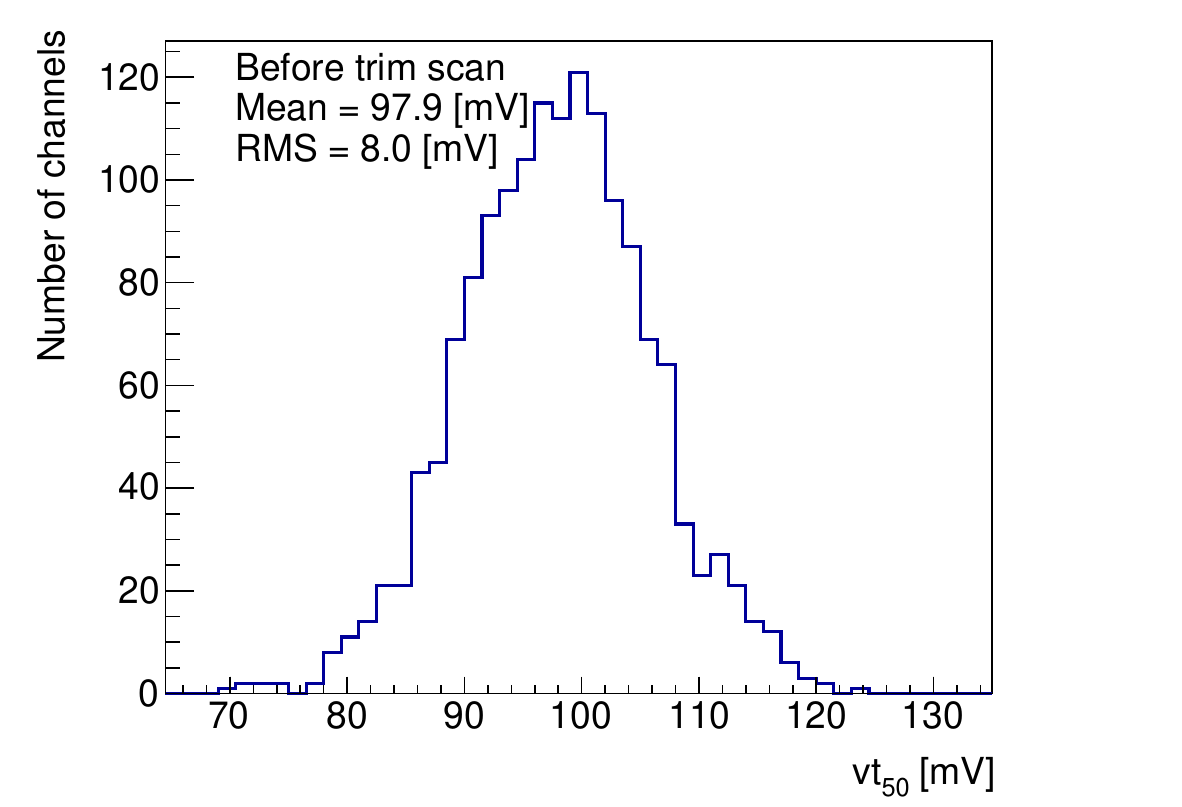}}
\subfloat[\label{f:trim_after_vt50}]{\includegraphics[width=0.49\textwidth]{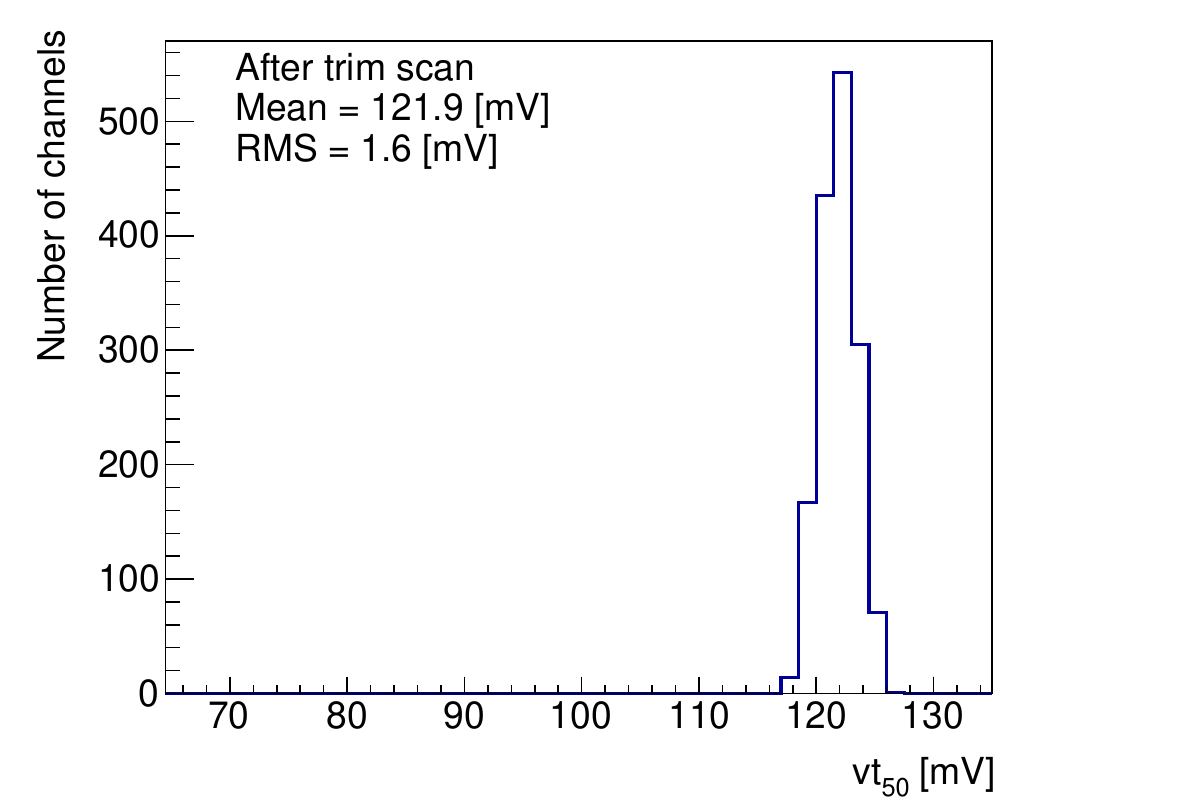}}
\caption{Threshold scan for an example module with 1.5\,fC injected charge before (a), and after (b) optimal trim settings from the trim scan are applied. The distributions of the $vt_\text{50}$ points for all strips are also shown before (c), and after (d) the trim scan.}
\label{f:trim}
\end{figure}

After the trim scan, an even response between different channels in a module is obtained as shown in Fig.~\ref{f:trim}. The majority ($99.9\%$) of the channels can be trimmed using the two lowest trim range settings, as expected for unirradiated modules, and an additional $0.05\%$ of channels can be trimmed using one of the larger trim ranges. The remaining channels cannot be trimmed even at the largest trim range setting.

\begin{figure}[]
\centering
\subfloat[\label{f:noise}]{\includegraphics[width=0.49\textwidth]{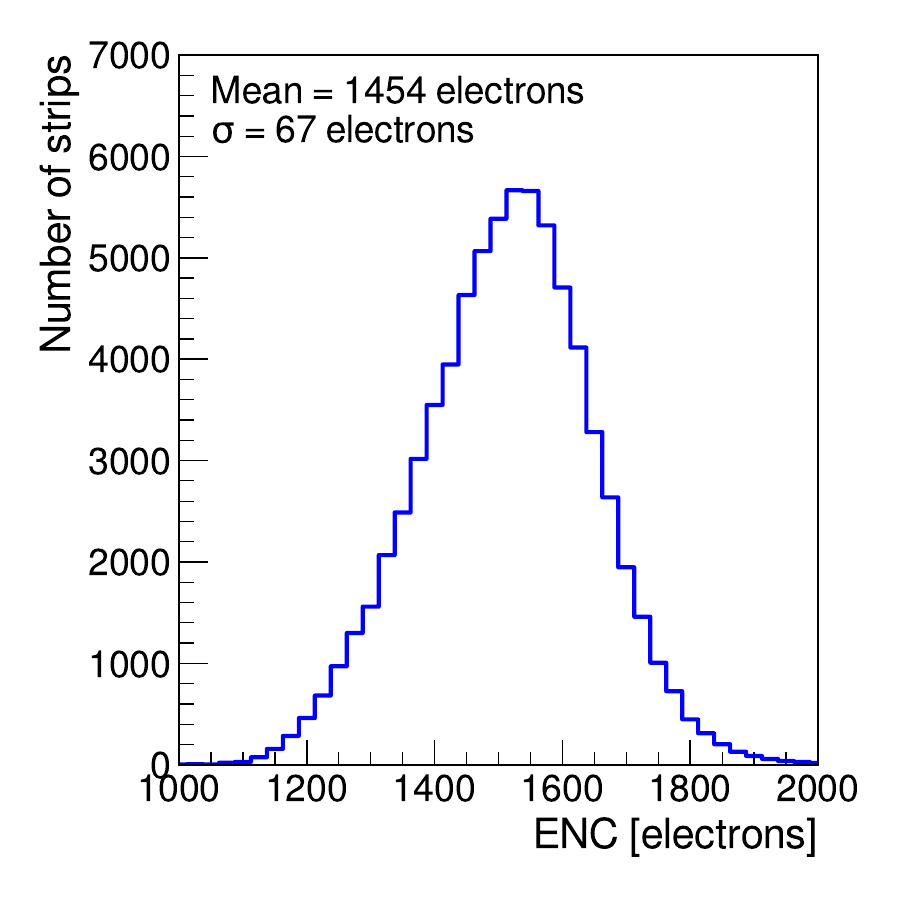}}
\subfloat[\label{f:noise-direct-no}]{\includegraphics[width=0.49\textwidth]{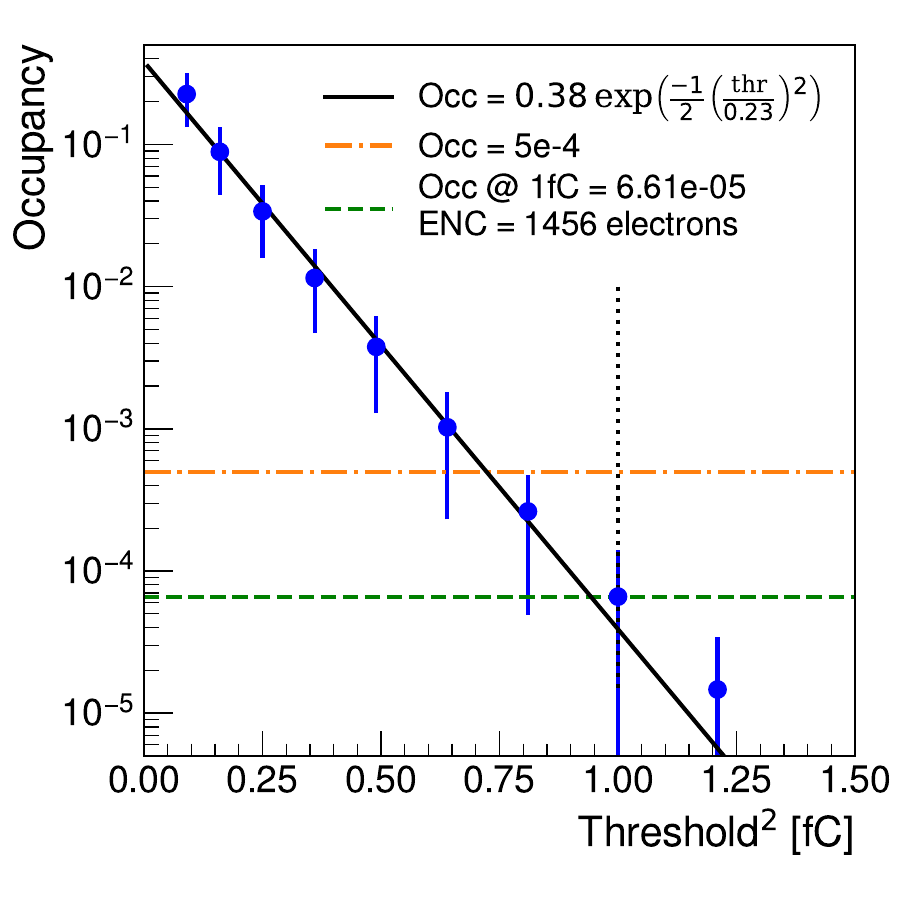}}\\
\subfloat[\label{f:three-point-gain_noise_occ_correclation}]{\includegraphics[width=0.53\textwidth]{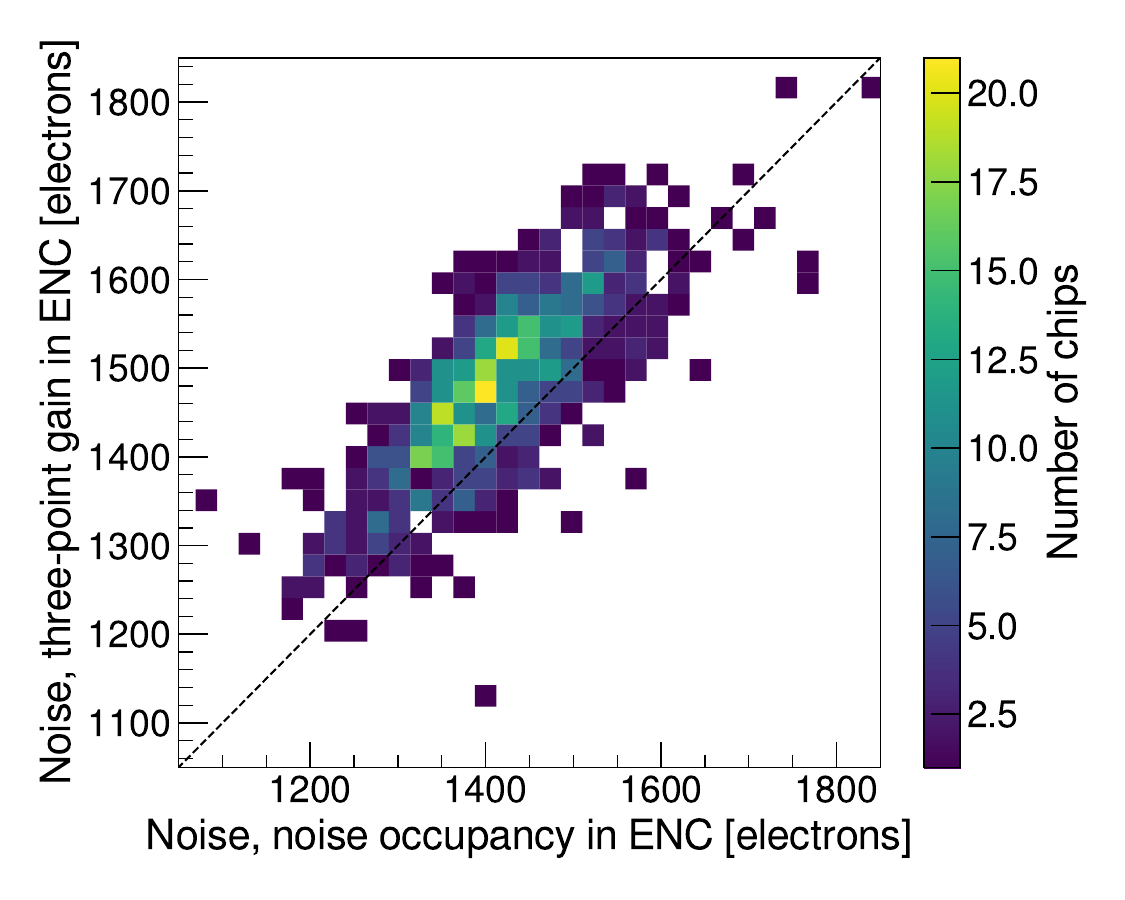}}
\subfloat[\label{f:strip-occupancy}]{\includegraphics[width=0.42\textwidth]{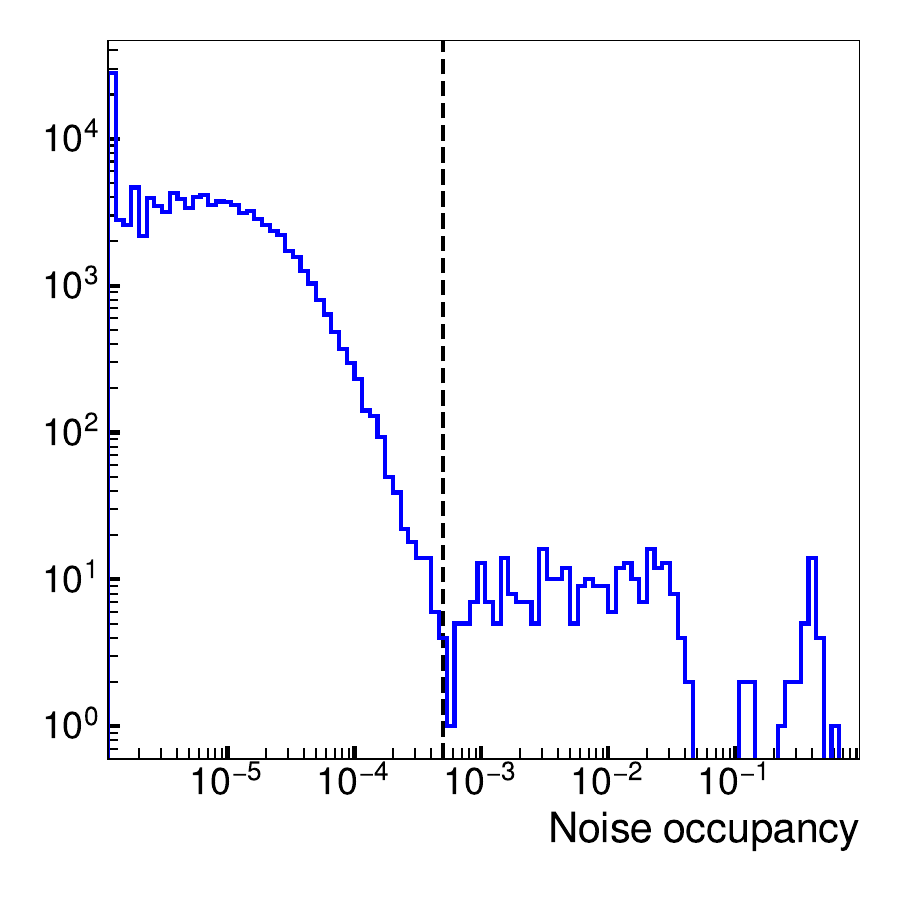}}
\caption{(a) ENC of all strips in the three tracker stations, (b) noise occupancy scan for a single module, (c) correlation of chip-by-chip ENC for the three tracker stations measured in the three-point-gain and in the noise occupancy scans, and (d) strip-by-strip noise occupancy measured from random triggers during combined runs. The dashed vertical line indicates noise occupancy of $5\times10^{-4}$.}
\label{f:noise-summary}
\end{figure}

It is important to keep the level of readout noise as low as possible to maintain low thresholds and accordingly realize high tracking efficiency.
An estimate of the ENC at the discriminator input is obtained, using the three-point-gain measurement during the routine calibration scans described in Section~\ref{sec:calibration}. This provides a measurement of the output noise value with an injection charge of 2\,fC. The distribution of ENC for all strips in the three stations are shown in Fig.~\ref{f:noise}, whose mean value is $1454 \pm 67$\,electrons. This is in good agreement with the $\sim 1500$\,electrons expected for unirradiated modules from the specifications~\cite{Abdesselam:2006wt}.

The additional dedicated noise occupancy scans were also performed for a more direct measurement of the noise. The results for a single module are shown in \cref{f:noise-direct-no}. Due to the large number of triggers required at high thresholds these take a considerably longer time than the three-point-gain measurement. The noise occupancy scans are, however, more sensitive to the tails of the noise distribution and to external sources such as a commonly increased noise on several channels (common-mode noise). The ENC from the noise occupancy scans is extracted from a linear fit to log (noise occupancy) vs threshold$^{2}$ in fC$^{2}$. The fit function used is $\mathrm{NoiseOccupancy} = N e^{- \frac{1}{2}\left(\frac{\mathrm{threshold}}{\sigma_{\mathrm{thr}}}\right)^{2}}$ where $N$ and $\sigma_{\mathrm{thr}}$ denote the normalization and ENC in fC, respectively. The chip-by-chip ENC evaluated from the noise occupancy scan and the three-point-gain measurement correlate well as shown in \cref{f:three-point-gain_noise_occ_correclation}, and are in good agreement with measurements of the ATLAS SCT modules~\cite{Abdesselam:2006wt}. \cref{f:strip-occupancy} shows the noise occupancy at the nominal 1\,fC thresholds using randomly triggered events during combined system runs \emph{in situ}, which provides a further cross-check of the measured noise. Over 99.7\% of the strips in the tracker satisfy the performance criteria that the noise occupancy per channel is less than $5\times10^{-4}$ at the nominal 1\,fC threshold. The main reason for noisy strips is a non-linear behavior in the response curve.

\subsection{Magnetic field test}
One of the concerns for the installation of the FASER tracker in between the gaps of the permanent FASER magnets was the presence of the stray magnetic field during the lowering of the tracker planes.
In order to exclude any negative impact on the performance or damage due to electromagnetic induction a test stand was set up that allowed to lower the FASER prototype tracker plane vertically in a controlled way at different distances to the aperture of one of the FASER magnets for testing purposes.

In a first step, the tracker plane was positioned $1\,\mathrm{m}$ away from the aperture (reference position) and a standard calibration and characterization sequence was run, {\it i.e.} the identification of noisy and non-responsive strips as well as the measurement of gain, threshold, electronics noise, and the sensor IV response.
At the reference position the magnetic stray field is negligible.
In a second stage, the mechanical mounting structure was placed in a distance of 65~mm from the magnet aperture which corresponds roughly to the distance of the first tracker plane in the final FASER assembly.
At this location, the magnetic field strength of the stray field amounts to up to $60\,\mathrm{mT}$.
The plane was lowered at a very low speed of about $0.03\,\mathrm{m}/\mathrm{s}$ until it was finally centered in front of the magnet aperture.
During the lowering process all cables (data, power, DCS) remained attached to the patch-panel of the plane, but were disconnected at the off-detector end.
A second set of performance measurements was taken at this position.
In a final step, a further reference measurement was taken at the reference position after the plane has been extracted in the reverse order with respect to the installation.

The number of noisy and non-responsive strips was found to be very consistent between the different measurement positions (maximally 2 strips of the difference in the prototype tracker plane).
The same conclusion applies to the measured gain and noise, {\it{i.e.}}, maximally $\pm0.3$\% and $\pm4$\% difference in the gain and noise of each module, respectively.

In conclusion, no sign for any damage or performance degradation could be found in this installation tests.

\subsection{Cosmic ray test}
In order to test the FASER tracker station as a full detector system including trigger and DAQ, a cosmic ray test stand was set up in the FASER surface laboratory.
Figure~\ref{f:setup} shows the setup in which the tracker station is placed horizontally in between two large trigger scintillators covering the full detector acceptance.

\begin{figure}[th]
\centering
\includegraphics[height=0.6\textwidth]{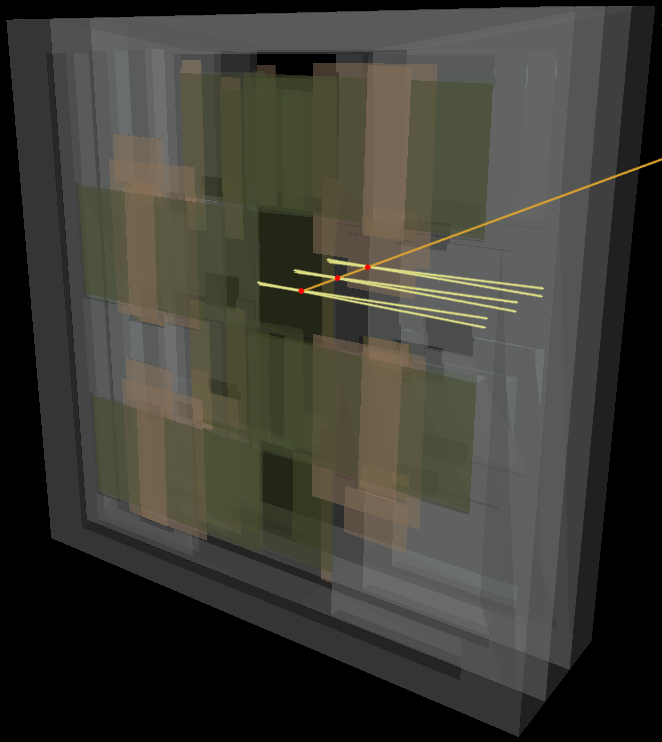}
\caption{Event display of a cosmic ray candidate traversing all three layers of one FASER tracker station, showing the hit strips (yellow), reconstructed space points (red) and tracks (orange).}
\label{f:cosmics}
\end{figure}

Figure~\ref{f:cosmics} depicts an example event display of a cosmic ray candidate traversing all three tracker layers leaving a clear signature.
The reconstructed space points from the crossing of two strips per layer are well compatible with a  straight track through the detector.

Beyond the bare proof of the tracker functionality and its interplay with the larger FASER trigger/DAQ system, the collected cosmics data set is very valuable for the intra-station alignment in the offline reconstruction due to the absence of any stray magnetic field which will not be the case once integrated in between the FASER permanent magnets.
The successful cosmic ray test concluded the tracker station commissioning on the surface.

\subsection{Summary of commissioning}

Three stations consisting of nine planes were successfully assembled and commissioned on the surface and subsequently \emph{in situ} after installation. Overall a very good performance was found, which is comparable between surface plane commissioning and station commissioning and in the tunnel.
The long term stability of full stations were additionally confirmed in long runs which lasted several weeks and was partially accompanied with data taking of cosmic ray events.

\section{Conclusion and outlook}
\label{sec:conclusion}

The FASER spectrometer was constructed with three tracker stations, comprising of nine planes of the silicon strip modules that were originally the spares for the ATLAS SCT barrel detector. All other parts of the tracker, including dedicated support frames, DCS and cooling system, readout electronics and services were newly developed for FASER.
To ensure a precise alignment, within the mechanical tolerances, after the construction and assembly metrology of the tracker was performed for each plane and station.

The installed detector is fully operational and shows excellent performance, well within the specifications. The number of dead channels is less than $\sim0.3$\% (including the modules in the outer region). The electrical performance of the detector was tested after calibration of the ABCD3TA readout chips at each stage of the construction. During the \emph{in situ} commissioning of the final FASER setup, the average noise (ENC) was evaluated to be 1454 electrons, and over 99.7\% of the the strips in the tracker satisfy the requirement of the noise hit occupancy $<5 \times 10^{-4}$ at 1~fC threshold.
Long-term operation as part of the commissioning showed that all the SCT modules can be kept below the required maximum temperature of 35~$^{\circ}$C with coolant temperature of 15~$^{\circ}$C. In addition to standalone tests, the tracking detector was tested as part of the full FASER detector commissioning using the final TDAQ system, for example for combined cosmic ray data taking.

A new tracker station, identical to the three already installed in FASER, and called the “Interface tracker”, will be installed between the, yet to be installed, emulsion detector and the FASER spectrometer. This Interface tracker will allow matching tracks between the emulsion detector and the FASER tracker, and enable to distinguish $\nu_{\mu}$ and $\overline{\nu}_{\mu}$ interactions by measuring the charge of the produced muon in the FASER spectrometer. The Interface tracker construction and surface commissioning started in June in 2021, and the installation is planned for late-2021.

The FASER experiment will start physics data-taking in proton-proton collisions from the start of LHC Run 3 operations in 2022. The tracker will act as one of the crucial detectors to allow to search for new light long-lived neutral particles in FASER.

%\acknowledgments
\section*{Acknowledgement}
We thank the technical and administrative staff members at all FASER institutions, including CERN, for their contributions to the success of the FASER effort. 
We are grateful to the ATLAS SCT Collaboration for donating spare SCT modules to FASER, and to the LHCb Collaboration for the loan of the calorimeter modules.
We thank Steve Wotton and Floris Keizer for providing their single module readout system, which enabled us to evaluate the spare SCT modules. 
We gratefully acknowledge input from Steve McMahon for his careful reading of the FASER "Letter Of Intent" and his input on the initial design of the FASER tracking detector, and from Dave Robinson, Bruce Gallop, Nobu Unno and Taka Kondo for their very useful advice when we faced unexpected behavior of the SCT modules.  
In addition, we thank the CERN cooling and ventilation group (EN-CV) for their work related to the FASER cooling system. 
This work is supported in part by Heising-Simons Foundation Grant Nos. 2018-1135, 2019-1179,  2020-1840, Simons Foundation Grant No. 623683, U.S. National Science Foundation Grant Nos. PHY-2111427, PHY-2110929, and PHY-2110648, JSPS KAKENHI Grants Nos. JP19H01909, JP20K23373, JP20H01919, JP20K04004, and JP21H00082, by the Swiss National Science Foundation, and by Tsinghua University Initiative Scientific Research Program.

\bibliography{references}

\end{document}